\documentclass[referee, pdflatex,sn-mathphys-num]{sn-jnl}% Math and Physical Sciences Numbered Reference Style 

\usepackage{graphicx}%
\usepackage{multirow}%
\usepackage{amsmath,amssymb,amsfonts}%
\usepackage{amsthm}%
\usepackage{mathrsfs}%
\usepackage[title]{appendix}%
\usepackage{xcolor}%
\usepackage{textcomp}%
\usepackage{manyfoot}%
\usepackage{booktabs}%
\usepackage{algorithm}%
\usepackage{algorithmicx}%
\usepackage{algpseudocode}%
\usepackage{listings}%
\begin{document}

\title[Article Title]{Climate2Energy: a framework to consistently include climate change into energy system modeling}

\author[1,2]{\fnm{Jan} \sur{Wohland}}%\email{jan.wohland@its.uio.no}
\author[2]{\fnm{Luna} \sur{Bloin-Wibe}}
\author[2]{\fnm{Erich} \sur{Fischer}}
\author[3,4]{\fnm{Leonhard} \sur{Göke}}
\author[2]{\fnm{Reto} \sur{Knutti}}
\author[4]{\fnm{Francesco} \sur{De Marco}}
\author[2]{\fnm{Urs} \sur{Beyerle}}
\author[5]{\fnm{Jonas} \sur{Savelsberg}}

\affil[1]{Department of Technology Systems, University of Oslo, Kjeller, Norway}

\affil[2]{Institute for Atmospheric and Climate Science, ETH Zurich, Zurich, Switzerland}

\affil[3]{Energy and Process Systems Engineering, ETH Zürich, Zürich, Switzerland}

\affil[4]{Reliability and Risk Engineering, ETH Zurich, Zurich, Switzerland}

\affil[5]{Energy Science Center, ETH Zurich, Zurich, Switzerland}

%%==================================%%
%% Sample for unstructured abstract %%
%%==================================%%

\abstract{
Supply and demand in future energy systems depend on the weather. 
We therefore need to quantify how climate change and variability impact energy systems. 
Here, we present Climate2Energy (C2E), a framework to consistently convert climate model outputs into energy system model inputs, covering all relevant types of renewable generation and demand for heating and cooling. 
C2E performs bias correction, uses established open-source tools where possible, and provides outputs tailored to energy system models. 
Moreover, C2E introduces a new hydropower model based on river discharge. 
We analyze dedicated hourly CESM2 Climate Model Simulations under the SSP3-7.0 scenario in Europe, covering climate variability through multiple realizations. 
We find large reductions in heating demand (-10\% to -50\%) and Southern European hydropower potentials (-10\% to -40\%) and increases in cooling demand (\textgreater 100\%). 
Based on stochastic optimizations with AnyMOD, we confirm that energy systems are highly sensitive to climate conditions, particularly on the demand side. 
}

%\keywords{renewable energy, climate change, decarbonization, climate risk}

\maketitle

\clearpage

\section{Introduction}\label{sec:Intro}

% Climate+Energy is important and urgent
While climate and energy system modeling emerged separately, they must be combined to tackle current energy system challenges for two main reasons.
First, the transition to renewable energy sources such as photovoltaic, wind, and hydropower makes supply sensitive to changes in weather and climate \cite{tobin_climate_2016, jerez_impact_2015, karnauskas_southward_2018, hou_climate_2021, losada_carreno_potential_2020, bloomfield_quantifying_2021-1, wohland_more_2017, pryor_climate_2020, van_vliet_power-generation_2016}. 
Moreover, the electrification of heating and fast uptake of space cooling increase demand sensitivity to weather and climate \cite{auffhammer_climate_2017, wenz_northsouth_2017, staffell_global_2023, bloomfield_quantifying_2021, iea_electricity_2025}. 
Second, climate change is a reality \citep{ipcc_climate_2021} and changes the operating conditions for energy systems. 
The energy system implications of climate change must therefore be considered.

% There are good open tools but it is unclear how to use them with CC
Combining climate and energy modeling is a complex interdisciplinary challenge \cite{craig_overcoming_2022}.
While high-quality open-source conversion tools exist for renewable generation potential \citep{pfenninger_long-term_2016, staffell_using_2016, haas_windpowerlib_2019, hofmann_atlite_2021} and weather-dependent energy demand \citep{staffell_global_2023, ruhnau_time_2019}, those tools are designed for use with historical weather data. 
Using them meaningfully with climate change projections necessitates, for example, relevant bias-corrected inputs and an assessment of climate variability. 

% CMIP and CORDEX are lacking bc they do not provide the right variables on the right resolution
Published studies generally draw from databases that were not designed to be energy system model inputs (CMIP \cite{eyring_overview_2016} or CORDEX \cite{jacob_euro-cordex_2014}), and available variables are therefore not tailored to energy system modeling needs.
The choice of variables and resolutions is consequently dominated by what is available as opposed to what would be desirable. 
For instance, near-surface winds are generally available while hub height winds are not, and most studies use vertical extrapolation heuristics despite known drawbacks \cite{hahmann_current_2022, wohland_extrapolation_2023}. 
Similarly, hydropower generation is typically modeled using runoff \cite{van_der_most_extreme_2022, gotske_future_2021}, creating a need for river routing assumptions in the hydropower conversion even though the climate models could provide internally consistent river discharge.
Moreover, desired and available resolutions typically diverge.
While many energy system models expect hourly inputs to capture the diurnal cycle (e.g., PyPSA-EUR \cite{horsch_pypsa-eur_2018}, Calliope \cite{pfenninger_calliope_2018}, Nexus-E \cite{gjorgiev_nexus-e_2022}), climate models usually provide coarser data. 
While this resolution gap can be narrowed down with regional climate models (e.g., EURO-CORDEX \cite{jacob_regional_2020}), important inconsistencies exist in aerosol and land use change implementations between EURO-CORDEX and CMIP which cause conflicting estimates of changes in future wind and solar energy potentials \cite{boe_large_2020, gutierrez_future_2020, wohland_process-based_2022}.
Regional climate model output is therefore not a silver bullet that provides the exact data needed. 

% PECD not a silver bullet either
The new pan-Euopean climate database (PECD) \cite{dubus_towards_2022} provides estimates for renewable energy potentials up to 2065 to be used, for instance, by transmission system operators.
While this update represents a major improvement, the methodology remains closed, making replication, modification and extension impossible. 
Moreover, PECD draws from existing climate data sets, and thereby inherits the limitation discussed above. 
In addition, it does not cover changes in heating or cooling demand. 

% Existing studies barely cover the full climate - to - energy chain
Energy systems are spatially interconnected and combine different types of generation, as well as storage, to meet demand.
Changes in one technology could thus be amplified, offset or reversed by changes in another, calling for holistic approaches. 
However, many climate change impact assessments for the energy domain focus on single technologies, like wind energy \cite[e.g.,][]{tobin_climate_2016, karnauskas_southward_2018, wohland_more_2017} or solar energy \citep{jerez_impact_2015, muller_cmip-5_2019-1, hou_climate_2021, shi_climate_2024}.  
Progress towards holistic approaches requires a formalized open framework spanning from climate model output to energy system model input, allowing others to reproduce, alter, and ultimately improve upon earlier work.

\section{Methods}\label{sec:Methods}
\begin{figure}[ht!]
    \centering
    \includegraphics[width=\textwidth]{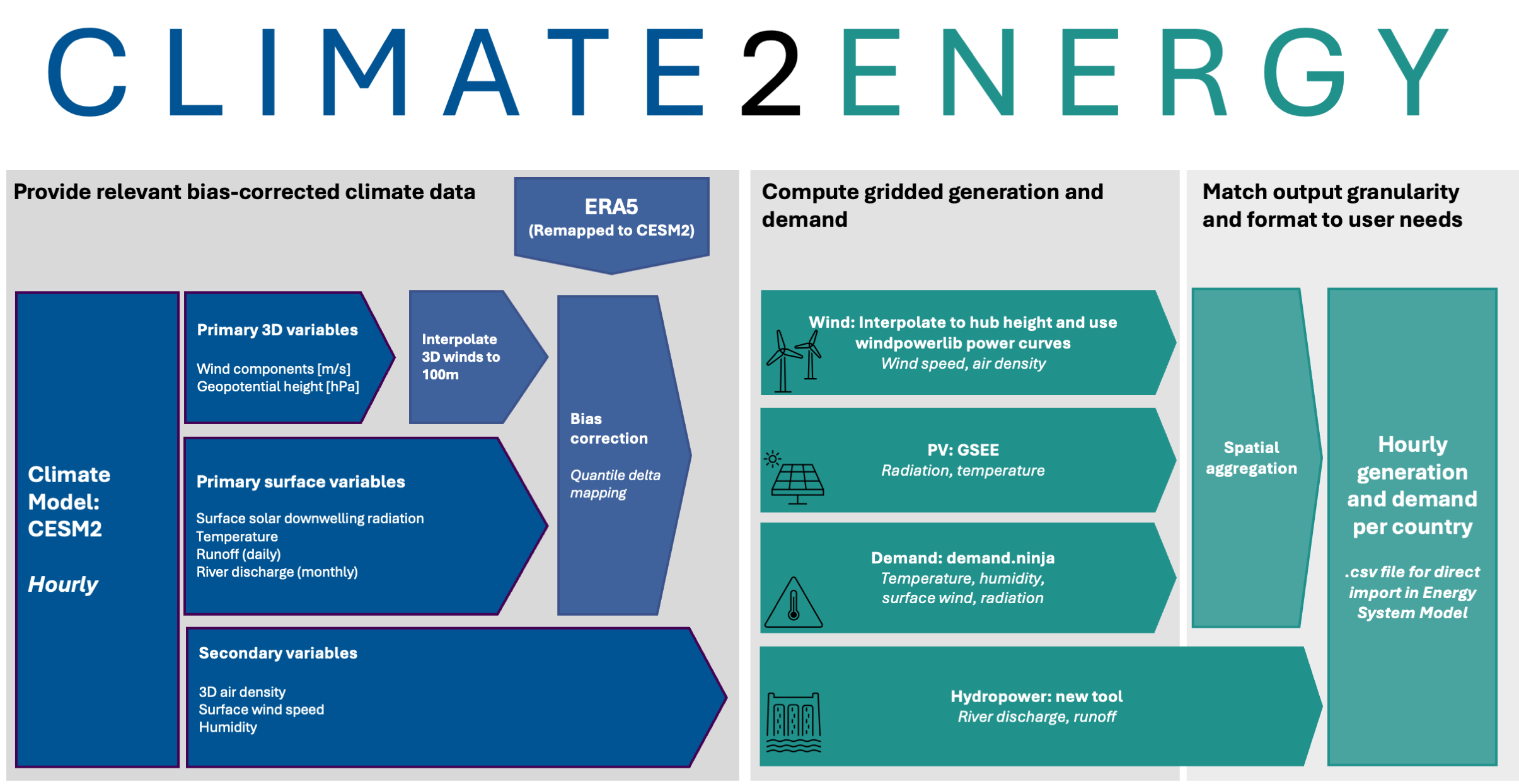}
    \caption{The Climate2Energy (C2E) framework to convert tailored climate model output to energy system model input. Key steps are separated by color: provision of relevant bias-corrected climate data (blue), computation of gridded generation and demand estimates (green), and postprocessing to match the desired data granularity (green). 
    %We run CESM2 to provide the required hourly atmospheric and land surface variables, followed by bias-correction with ERA5. 
    Where possible, we use established tools, for example, to compute PV and wind capacity factors, and heating/cooling demand. 
    We develop a new sub-module to compute hydropower generation and test it relative to existing databases. 
    Ultimately, C2E provides csv files on country granularity that can be immediately used in energy system optimization models.}
    \label{fig:flowchart}
\end{figure}

This study tackles three main challenges towards better energy-climate assessments. 
First, we provide tailored global hourly climate model output for the climate-to-energy conversion, showing that providing such data is straightforward for climate modeling groups without prohibitive data storage despite the hourly resolution and archiving of the 3D atmosphere. 
%Moreover, we perform three climate model realizations sampling European climate variability, allowing us to investigate the role of internal variability as well as climate change for energy-system applications in an unprecedented manner. 
Second, we develop and publish Climate2Energy (C2E), an open framework to translate climate model output into energy system model input, accounting for climate model biases and leveraging existing peer-reviewed tools to cover all relevant types of renewable generation, as well as heating and cooling demand (Fig. \ref{fig:flowchart}, \citep{staffell_global_2023, staffell_using_2016, haas_windpowerlib_2019, pfenninger_long-term_2016-1}).
Third, we evaluate the energy-system implications of climate change with a state-of-the-art stochastic power system model.
%Taken together, we present a concrete framework of how energy-climate modeling can be taken to the next level. 

\textbf{Climate model, scenario, and realizations}

We run the climate model CESM \cite{danabasoglu_community_2020} in version 2.1.2 with hourly outputs that are tailored to the energy conversion (see Table \ref{tab:variable_overview}).
Some variables are standard output but normally unavailable at hourly resolution (RSDS, $z_g$, $T$, $Q$, $q$, $U10$), while others are generally not stored ($\rho$, $U$, $V$). 

\begin{table}
    \centering
    \begin{tabular}{c|c|c|c}
         \textbf{Variable name} & \textbf{Abbreviation} & \textbf{3D/2D} & \textbf{temporal resolution} \\
         \hline
         Wind components & $U$, $V$ & 3D & hourly\\
         Air density & $\rho$ & 3D & hourly \\
         Geopotential height & $Z_g$ & 3D & hourly \\
         Surface solar downwelling radiation & RSDS & 2D & hourly\\
         Surface temperature & $T$ & 2D & hourly \\
         Runoff & $R$ & 2D & hourly\\
         River discharge & $Q$ & 2D & monthly  \\
         Specific humidity & $q$ & 2D  & hourly \\
         Surface wind speeds &$U_{10}$ & 2D & hourly\\
    \end{tabular}
    \caption{Overview of the tailored CESM2 climate model output. 3D variables are stored on model levels.}
    \label{tab:variable_overview}
\end{table}

To compare climate change to internal climate variability, we provide three realizations which are branched from an existing 11-member ensemble at the beginning of the historical (1995-2015) and future (2080-2100; SSP3-7.0) periods.
We seek to maximize variability in our sample by choosing two realizations based on the phase of the North Atlantic Oscillation (NAO) index in the respective period (most positive and most negative).
We choose the third ensemble member randomly. 

\textbf{Bias correction}

Climate models exhibit biases relative to observations and reanalyses, and are therefore usually bias-corrected in climate impact studies \citep{maraun_bias_2016}.
We perform bias correction \textit{before} the climate-to-energy conversion because of non-linearities in the conversion, using univariate delta quantile mapping per grid box to correct CESM2 biases relative to ERA5 over the 1995-2015 period.
This approach leaves the spatial and inter-variable rank correlations unchanged and is therefore well suited for energy system studies where balancing across technologies and locations are key strategies to mitigate variability. 
We assume bias stationarity in the future, as is commonly done.
We bias correct all primary variables (RSDS, $T$, 100m wind speed, and $Q$) and use the secondary variables ($q$, $U_{10}$, $\rho$) without bias correction to limit computational complexity and because they only marginally impact the outputs. 

Each bias-corrected future realization is tied to a historical realization via the bias correction.
Since the choice of the historical realization matters \citep{Bonnet_Sensitivity_2022}, we use all possible combinations, yielding 9 future realizations (3 plausible bias corrections x 3 future scenarios). 
It is justified to combine future realizations with different historical realizations because the periods are separated by 65 years, and atmospheric variables do not exhibit relevant memory on these timescales. 

\textbf{Solar Energy}

The conversion to solar capacity factors (CFs) uses the Global Solar Energy Estimator \cite{pfenninger_long-term_2016}. 
Solar PV generation mainly depends on surface radiation while also being impacted by surface winds and temperature via efficiency changes. 
We assume Silicon-made panels that point south to maximize yield with a tilt of 35$^\circ$. 

When aggregating solar and wind CFs to country level, we only consider sites that are better than the median on average because infrastructure siting favors resource-rich locations \citep{monforti_how_2016}. 

\textbf{Wind Energy}

The wind CF conversion uses model-level winds to avoid problematic vertical extrapolation \citep{wohland_mitigating_2021, wohland_extrapolation_2023}. 
Specifically, we interpolate from adjacent model levels to 100m, bias correct the 100m wind speed, and then interpolate to hub height. 
The intermediate step at 100m is necessary because ERA5 winds, used for bias correction, are available at that height. 
We interpolate by fitting a power law with a temporally and spatially evolving exponent $\alpha(\Vec{x},t)$, see \citep{wohland_extrapolation_2023}. 
We analyze the same turbines as in \citet{wohland_mitigating_2021}, namely SWT142\_3150 (129m hub height), SWT120\_3600 (90m hub height), and E-126\_7580 (127m hub height).
Turbine power curves are from windpowerlib \citep{haas_windpowerlib_2019}.

Air density is relevant because wind energy density is proportional to it.
We include air density changes via wind speed modification to avoid implausible changes of the rated capacity, following  \citep{svenningsen_power_2010}, see SI \ref{app:density} for details. 

\textbf{Hydropower}

We convert CESM2 river discharge into run-of-river generation and reservoir inflow using a piecewise linear regression calibrated on historical river discharge \citep{harrigan_glofas-era5_2020} and ENTSO-E hydropower production \citep{entso-e_entso-e_nodate, hirth_entso-e_2018}.
This approach captures that hydropower generation generally scales with water throughput, except when operational bounds are crossed and water spillover occurs. 
We include those countries with the largest hydropower production according to PECD \cite{de_felice_entso-e_2022}, covering 83\% of European run-of-river generation, 80\% of reservoir inflows, and 99\% of pumped hydro inflows (details in SI Fig. \ref{fig:PECD_hydro_share}). 
Since new hydropower projects are rare, we use currently installed capacities in our conversion. 

Our approach ensures good agreement of the CESM2-based hydropower estimates with observations in terms of annual means with mean relative errors of about 6\% (SI Fig. \ref{fig:an_prod}), as well as good rank correlation on a per timestep basis (generally $>0.7$, SI Table \ref{tab:corr}).
While the method also reproduces the ENTSO-E seasonal cycles well when driven with ERA5 inputs, results are varied when using CESM2 inputs (SI Fig. \ref{fig:season_cycle}).
Overall, we conclude that our hydropower approach is suitable for the analysis in this study, despite climate model limitations, and we refer to the SI Section \ref{sec:hydro_details} for a detailed discussion and extended documentation.

\textbf{Heating and cooling demand}

We use demand.ninja \citep{staffell_global_2023} to compute heating and cooling demand as a function of temperature, thermal inertia, wind chill, radiative gains and human temperature perception.  
Specifically, we use country-specific thresholds for the heating and cooling onset temperatures and electricity demand sensitivity from  \cite{staffell_global_2023}.
Data is unavailable for some small Southern European countries (Albania, Bosnia-Herzegovina, Croatia, Montenegro, Macedonia, Serbia, and Slovenia), with less than 25 million inhabitants.  
We compute their threshold as the mean over their neighbors, weighting by population in case of capacities. 

CESM2 provides all inputs that demand.ninja requires, except 2m winds, which we derive from 10m winds using a power law with $\alpha=1/7$. 

C2E outputs population-weighted country means in two different scenarios: (1) using the currently electrified share and (2) assuming full heating sector electrification.
To compute electricity demand for heating under full electrification, we scale outputs using the currently electrified heating share from JRC-IDEES \cite{rozsai_jrc-idees-2021_2024}. 
When countries are not covered in the JRC database, we assume that their electrified share equals that in similar countries, see Table \ref{tab:similar_countries}.

\begin{table}[h]
    \centering
    \begin{tabular}{c|c}
        \textbf{Country not covered} & \textbf{Similar country} \\ \hline
        Albania   & Romania   \\ 
        Bosnia and Herzegovina   & Slovakia  \\ 
        Macedonia   & Slovenia   \\ 
        Montenegro   & Slovakia   \\ 
        Serbia   & Bulgaria   \\ 
        Norway   & Sweden   \\ 
        Switzerland   & Austria   \\ 
    \end{tabular}
    \caption{Mapping between countries that are not covered in the JRC-IDEES database to similar countries that are covered. This mapping is only relevant for the \textit{fully-electrified} scenario.}
    \label{tab:similar_countries}
\end{table}

\textbf{Energy System Model}

To understand climate impacts on the energy system, we run a stochastic energy planning model based on the AnyMOD.jl framework using 20y of hourly C2E outputs. 
The model represents the energy system as a linear optimization problem deciding on investment and operation to minimize system costs while meeting constraints, like energy balances or emission targets \cite{goke_anymodjl_2021, goke_stabilized_2024}. 

We focus on the European power sector, covering dispatchable generation from gas turbines, long-term storage via electrolytic hydrogen, and battery storage, among others (overview in SI Fig. \ref{fig:energySystemModel}). 
The final electricity demand combines the weather-dependent fully-electrified C2E heating demand scenario with weather-insensitive remaining electricity demand from \cite{bourmaud_osmose_2022}.  
We cover the 32 European countries displayed in Fig. \ref{fig:mean_map}. 

Average annual carbon emissions are limited to 10 megaton which is very low compared to current emissions.
We thus essentially enforce carbon neutrality while avoiding that rare and unrepresentative periods with very low renewable generation blur out the effects of different weather inputs that are studied here. 

%allowing Europe to not be fully self-sufficient and solely rely on a fluctuating renewable supply, which could result in rare periods with low renewable supply disproportionately impacting system design. 
%Overfitting to these rare and unrepresentative periods would blur out the average effects of different weather inputs that we focus on here. 
%Instead, small emissions from flexible generators can be offset by negative emissions in other sectors or preferably avoided altogether if, instead of fossil fuels, renewable fuels, like hydrogen or synthetic methane, are imported and utilized.

The model optimizes across all operational years. %, and it does not consider a multi-year transition pathway. 
We fix hydropower capacities at today's levels because of negligible expansion potential.
While we allow unmet demand, it is very expensive (13,000 €/MWh), corresponding to the social opportunity costs of power outages \citep{ropke_development_2013}. 
The model computes a cost-efficient system considering all 20 weather years from C2E at an hourly resolution in the historical and future period, respectively.
Overall, the setup contains 519 investment variables,  145 million operational variables, 156 million constraints, and 544 million non-zero elements in the underlying matrix.
%This approach balances security and costs, ensuring a reliable but not over-engineered power system.

For reference, we also solve the model separately for each weather year and compute their convex hull, which is defined by the maximum for each of the 519 investment variables across all the model results. 
As a result, the convex hull guarantees a reliable system for the whole dataset, requiring an extensive dataset but no stochastic optimization. 

\section{Results and Discussion}\label{sec:EnergyModelInputs}
\subsubsection*{Changes in mean generation and demand}
\begin{figure}[ht!]
    \centering
    \includegraphics[width=\textwidth]{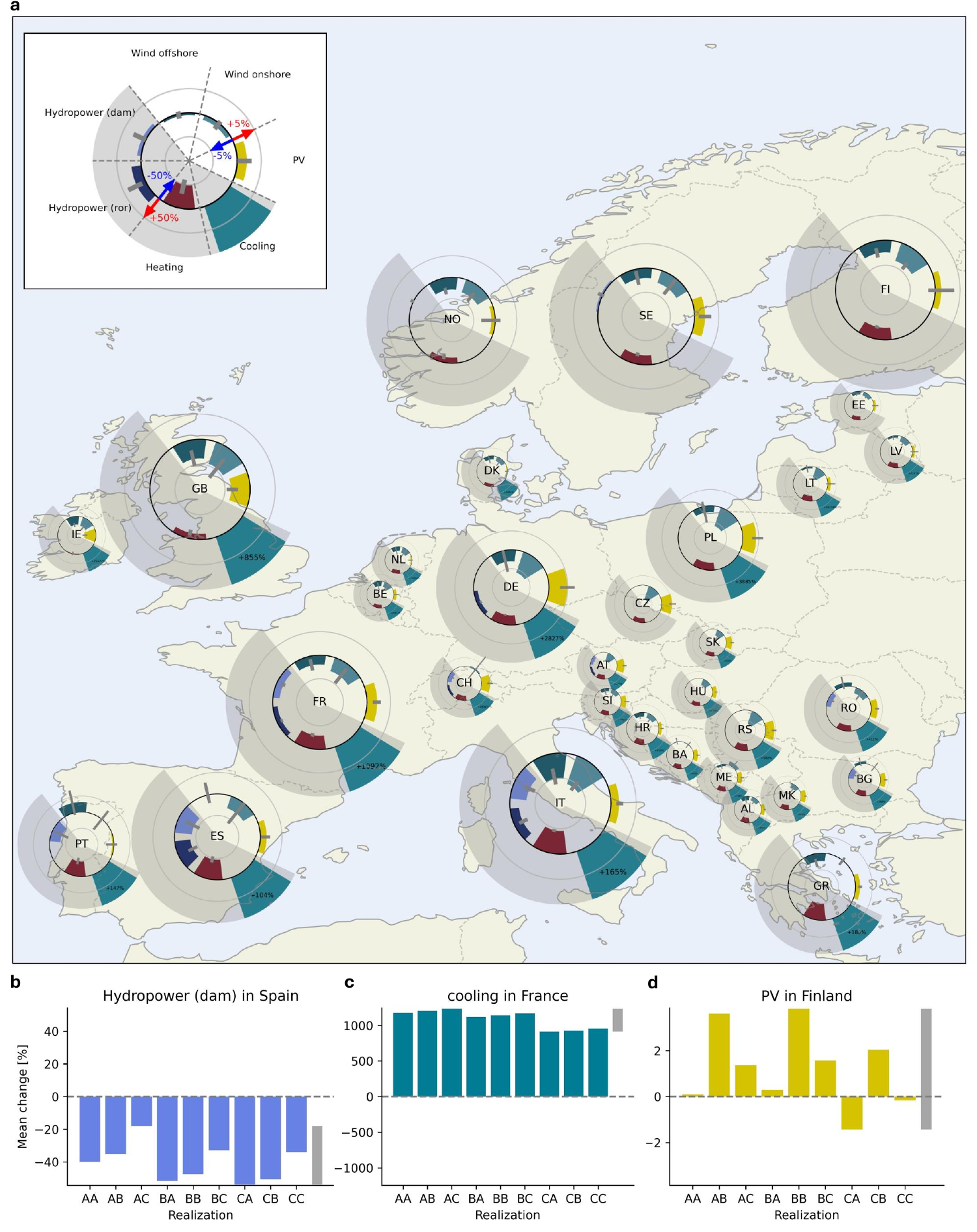}
    \caption{Mean changes in renewable generation and demand for heating and cooling in Europe in the SSP3-7.0 scenario (2080-2100 minus 1995-2015; in \%). Subplot a provides a continental overview covering all countries and technologies. It shows the ensemble mean and the max-min range of the underlying nine realizations as grey error bars. We use two different scales, as illustrated in the legend, because changes in hydropower and demand are larger than changes in PV and wind energy. Increases exceeding the scale are provided as annotations. Subplots b-d show examples of technology-country combinations for all realizations, and grey bars in in b-d correspond to grey bars in a.}
    \label{fig:mean_map}
\end{figure}

We find that average wind and solar CFs change moderately ($<5$\%) while hydropower, heating demand, and cooling demand change strongly (generally $> 10$\%) between the reference and future period under SSP3-7.0 (see Fig. \ref{fig:mean_map}a). 
Most changes are greater than what is expected from climate variability (gray uncertainty ranges in Fig. \ref{fig:mean_map}), and therefore represent systematic changes in mean conditions that future power systems must cope with. 

% Variable renewable generation

Both onshore and offshore wind capacity factors (CFs) decrease by a few percent in most countries, including in demand centers like Great Britain, France, Germany, and Italy. 
These reductions are a combination of dynamical effects, land surface changes and air density reductions (see SI  \ref{app:density}). 
Wind CF changes in Iberia and Greece, by contrast, remain dominated by climate variability and are therefore uncertain, in line with a previous study identifying this area as a hotspot of multidecadal wind generation variability \cite{wohland_mitigating_2021}. 

Solar CFs increase by a few percent in most countries with relatively strong increases of about 5\% in Central Europe. 
These increases occur despite reduced panel efficiencies due to higher future temperatures, suggesting that changes in cloud cover and clear-sky radiation dominate over efficiency losses.
Contrary to the general trend, PV CFs decrease by almost 5\% over the British Isles, and the sign of change is uncertain in North-eastern Europe. 
For example, PV CF changes in Finland range from a -1\% reduction to +4\% increase, depending on realization and leading to a low mean change with a large uncertainty (cf. Fig. \ref{fig:mean_map}d).

Those changes in wind and solar Cfs matter even though reductions of a few percent might appear small compared to uncertainties from technological progress or geopolitical risks.
They matter because they represent geophysical constraints and because some countries are impacted multiple times. 
Great Britain, for example, has to cope with about 4\% reductions in onshore wind, offshore wind and solar PV simultaneously. 

% Hydropower and demand
Hydropower suffers almost everywhere, in line with the IPCC summary that Mediterranean runoff decreases with high confidence \cite{ipcc_climate_2021}.
Spain is impacted the hardest with about 40\% reduction in run-off-river generation and inflows into reservoirs (cf. Fig. \ref{fig:mean_map}b).
Only Norwegian and Swedish hydropower will not be negatively affected. 
While Norway's mean reservoir inflows remain unchanged, changes in Sweden are small ($<$1\%) and uncertain.

Heating will require less energy across the continent, as expected when temperatures rise. 
Southern European countries benefit the most, and the Greek, Italian and Spanish heating demands will be cut in half.
The effect is weaker in Northern Europe.
Scandinavia, for example, only experiences relative reductions of -10\% to -20\%. 
However, since those countries have higher absolute heating demand, their absolute demand reductions are highly relevant. 
There is one outlier to the general trend: Ireland, where heating demand is essentially unchanged, presumably because of its mild and wet maritime winters.   

We find the most dramatic changes for cooling demand that strongly increases, exacerbating a trend that is already visible  \citep{iea_electricity_2025}. 
Cooling demand is projected to rise  by 100\% to 200\% in southern countries that already use air conditioning extensively today like Portugal, Spain, Italy, and Greece.
In countries that lie slightly further North, like France, cooling demand increases about 10-fold (cf. Fig. \ref{fig:mean_map}c).
Relative increases in some countries are very high because of low baselines (e.g., +6735\% in Austria).

In summary, we reveal important changes in mean energy generation and demand: modest reductions in wind potentials, slight increases in PV potentials, strongly reduced hydropower potential, strongly reduced heating demand and skyrocketing cooling demand. 
%These changes in the mean represent only one aspect of the challenge because future highly renewable power systems use variable weather-dependent supply to meet weather-dependent demand. 
%We therefore proceed with an evaluation of the seasonal cycle and inter-country generation complementarity.

\clearpage
\subsubsection*{Changes in seasonality of generation and demand}

\begin{figure}[ht]
    \centering
    \includegraphics[width=.495\textwidth]{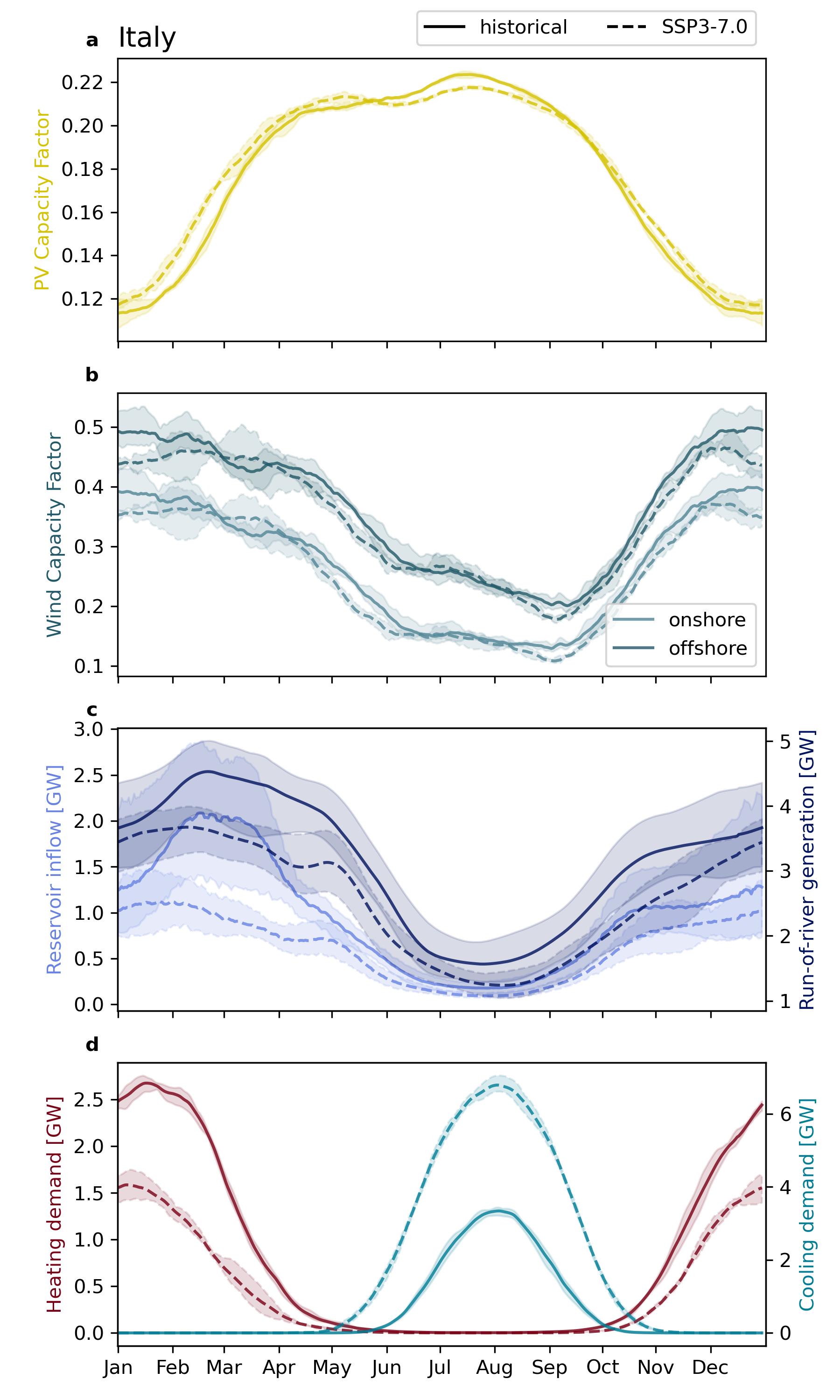}
    \includegraphics[width=.495\textwidth]{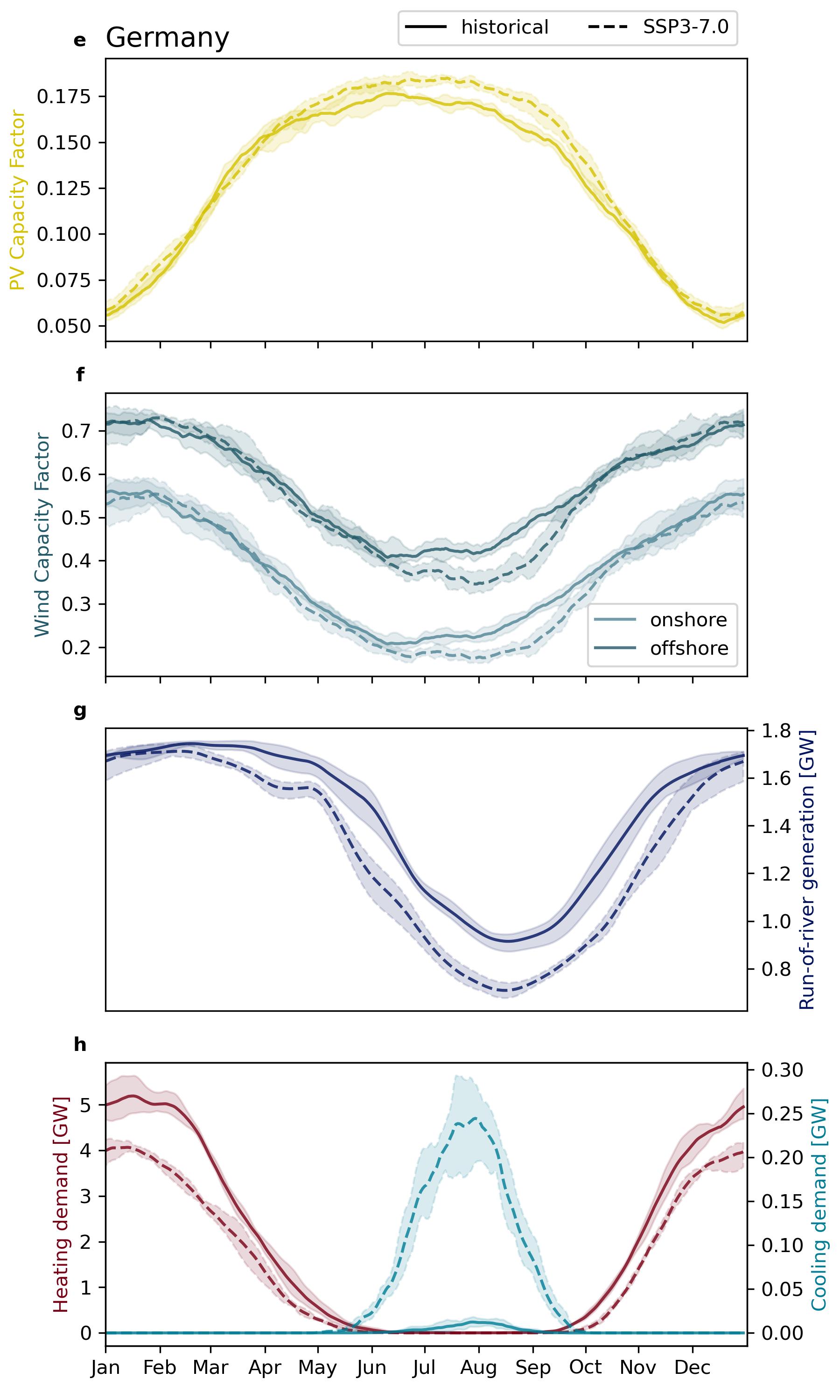}
    \caption{Historical and future seasonal cycles of renewable generation and heating/cooling demand in Italy (a-d) and Germany (e-h). Lines are averages over 20 years (1995-2015 historical; 2080-2100 SSP3-7.0) and nine realizations, smoothed using a 4 week running-mean. Colored areas show the min-max range of the realizations, providing insight into the relevance of internal variability. Onshore and offshore wind capacity factors are derived as the mean over the 3 considered wind turbines. Heating demand is reported for the currently electrified heating share.}
    \label{fig:seasonal_timeseries}
\end{figure}

Fig. \ref{fig:seasonal_timeseries} compares the historical and future seasonal evolution of generation and demand in two large countries (Italy and Germany).
%We pick those countries as examples and refer to the energy system modeling for a holistic assessment (Sec. \ref{sec:Results_energy}). 
While mean PV CFs increase slightly in both (about 5\% in Germany; 2\% in Italy), the timing differs. 
Italian PV is projected to generate more in winter and less in summer (Fig. \ref{fig:seasonal_timeseries}a), whereas German PV generation is projected to remain largely unchanged in winter and increase in extended summer (Fig. \ref{fig:seasonal_timeseries}e). 
While both countries thus show opposing seasonal cycle changes, the overall seasonal patterns (high PV generation in summer, high wind generation in winter) remain.
PV uncertainty bands are narrow, indicating a minor role of climate variability.

Changes in wind CFs, by contrast, are often unclear, indicated by overlapping uncertainty bands. 
For instance, future Italian capacity factors are lower between December and February but each curve is still within the uncertainty band of the other curve (Fig. \ref{fig:seasonal_timeseries}b).
Between July and October, however, German wind potentials are clearly reduced (Fig. \ref{fig:seasonal_timeseries}f). 
Since this reduction coincides with the period of increased PV CFs, it demonstrates the possibility to compensate for losses in wind energy with gains in solar PV.

Hydropower seasonality changes are also country specific.
They must be interpreted carefully because CESM has difficulties capturing hydropower seasonality (SI Sec. \ref{sec:hydro_details}).
Italian reservoir inflow and run-of-river generation both peak in spring during the historical period (Fig. \ref{fig:seasonal_timeseries}c).
This peak disappears in the future, presumably related to a shortened melting season in mountainous regions. 
By contrast, German historical and future run-of-river generation follow each other in January-February and then drop to their a minimum in about August. 
The already existing minimum becomes more pronounced and lasts longer in the future. 
That is, Italian hydropower mostly loses in spring, whereas German hydropower does so in extended summer. 

Heating and cooling demand projections indicate that heating demand shrinks considerably, and the heating season shortens. 
Cooling demand grows overall and the cooling season lengthens. 
The energy system implications of these expected changes are profound, as more demand must be met in summer and less in winter, increasing the value of PV, as reported for USA \citep{shi_climate_2024}. 
Moreover, the cooling demand changes exhibit strong regional variations. 
While future Italian cooling demand peaks at 6 GW, more than twice its currently electrified heating demand peak, the future German cooling demand peak is only about 5\% of currently electrified heat demand peak.

Taking into account the combination of the projected changes, it becomes evident that climate change impacts in different parts of the energy sector add up and exhibit strong seasonal variations. 
They can increase system stress, as in summer in Germany, where wind and hydropower generation drop while air conditioning drives up demand.
They can also offset system stress, at least partially, as in spring in Italy, where reduced hydropower generation coincides with reduced heating demand. 
In either case, however, a consistent climate-to-energy conversion covering all technologies is needed to capture the interconnectedness of energy sector climate change impacts.

\subsubsection*{Changes in spatial structure of generation and demand}

\begin{figure}[ht]
    \centering
    \includegraphics[width=.9\textwidth]{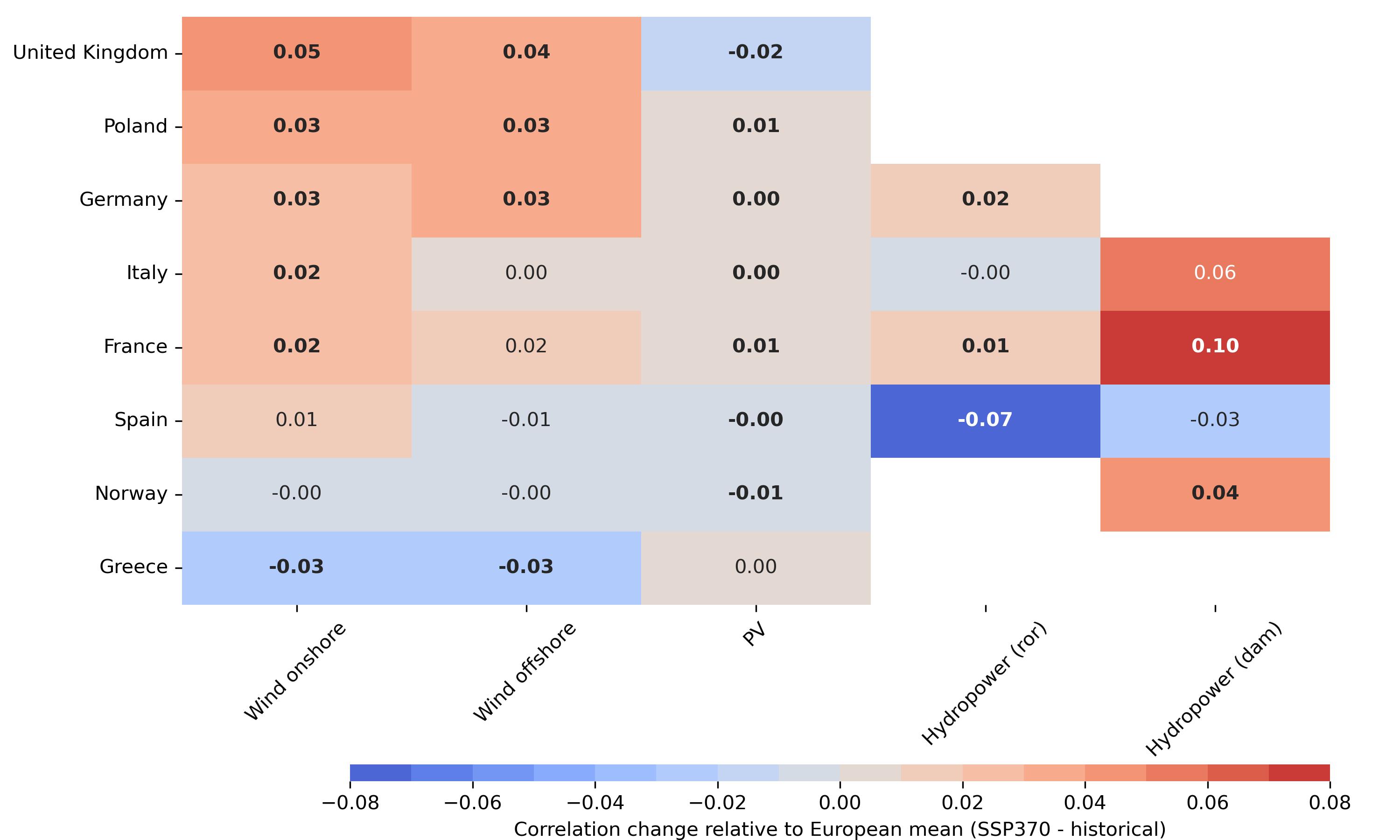}
    \caption{Changes in spatial complementarity of generation for a subset of covered countries. Values show Pearson correlation change between generation or demand in a given country and the European mean. Correlation is computed per realization and subsequently averaged. Bold font indicates that all 9 realizations agree on the sign of change. Red colors imply increased similarity of the timeseries in a given country and the European mean, and blue indicates the opposite. White boxes mean that a technology does not exist in a country because it has no coast (i.e., no offshore wind), is not covered in the databases used to calibrate the hydropower conversion, or because it currently has no cooling-temperature dependency according to demand.ninja. Countries are sorted by Wind onshore in descending order. A version of this table covering all countries is available in the SI Fig. \ref{fig:SI_correlation_change_all}.}
    \label{fig:correlation_change_subset}
\end{figure}

The complementarity of generation and demand between countries sets upper limits to the effectiveness of transmission infrastructure \citep{wohland_more_2017}. 
Fig. \ref{fig:correlation_change_subset} shows correlation changes between individual countries and the European mean.
Onshore wind energy correlations increase in most countries implying more similar generation timeseries and less potential for spatial balancing, confirming a previous study \citep{wohland_more_2017},.
We obtain similar, yet weaker and less consistent changes for offshore wind, presumably related to a larger overlap of the wind distribution with the turbine rated regime, where generation is insensitive to wind changes.
Greece deviates from the general trend: correlations decrease with full agreement across realizations, potentially due to a larger role of more local, thermal wind systems. 
PV correlations increase weakly and with high inter-realization agreement in most countries, in line with \citep{hou_climate_2021}. 

Hydropower dam inflows generally become more uniform, as shown by strong correlation increases in Italy, France, and Norway.
Run-of-river changes, however, are more varied: correlations strengthen mildly in Germany, and France, and weaken strongly in Spain, while being uncertain in Italy.
While these results indicate better complementarity between Spanish and European hydropower, they are caused by the stark deterioration of Spanish hydropower conditions and should not be over-interpreted (cf. Fig. \ref{fig:mean_map}a). 

\section*{Reduced capacity expansion needed to meet reduced demand} 
\label{sec:Results_energy}

\begin{figure}[ht]
    \centering
    \includegraphics[width=1.0\linewidth]{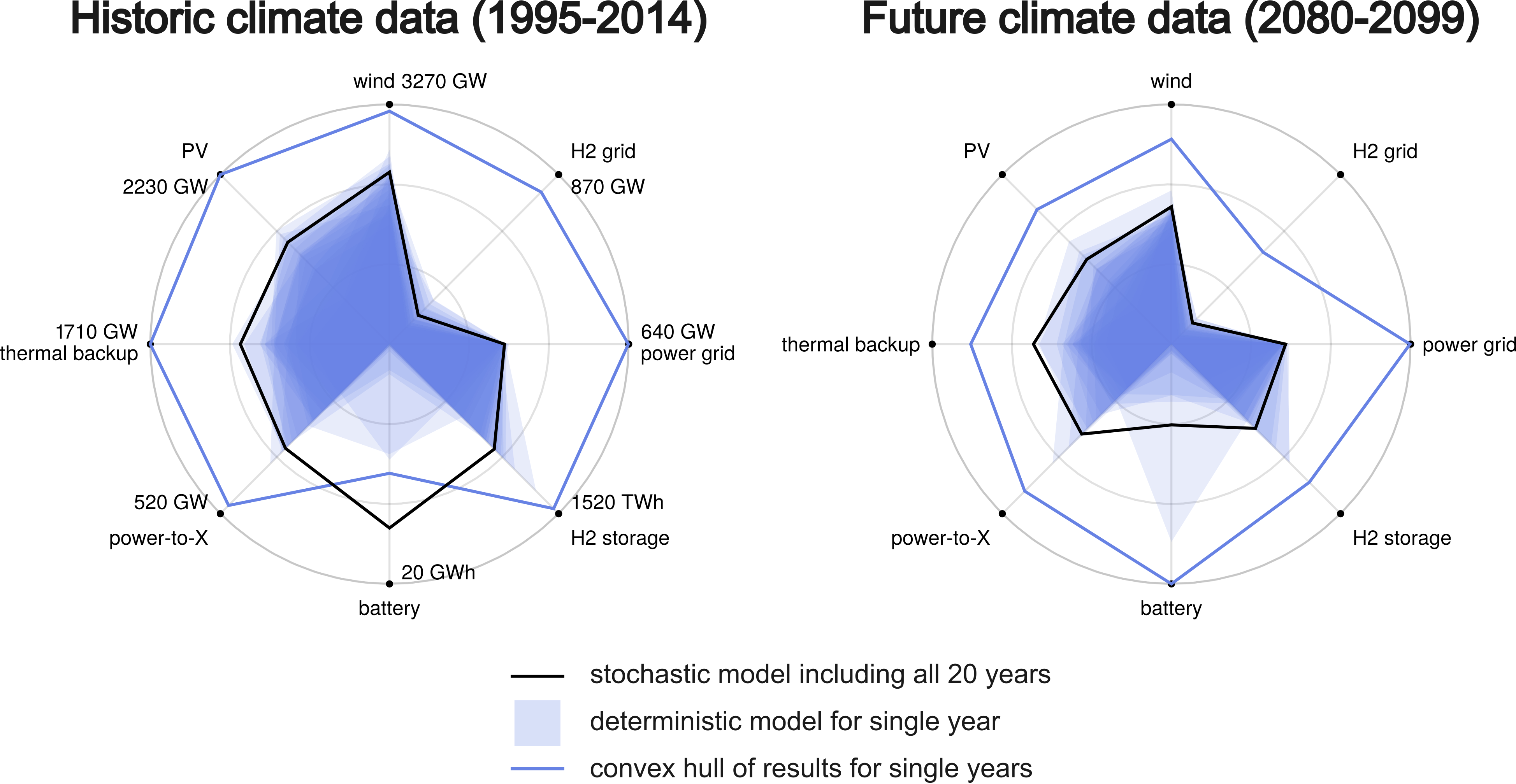}
    \caption{Aggregated European capacity layout using historical and future C2E inputs. The black line corresponds to one stochastic optimization over 20 years of hourly inputs simultaneously. Blue shapes correspond to separate optimization for one year at a time, and the blue line shows the maximum for each capacity variable across all 20 years (i.e., the convex hull).  The power-to-x category aggregates the two electrolyzer technologies and the methane pyrolysis; the thermal backup category all engines and open-cycle turbines. The scale and legend are the same for both radar plots. This plot is based on CESM2 realization AA, see SI Figs. \ref{si_fig:anymod_optimization_bb} and \ref{si_fig:anymod_optimization_cc} for other realizations.}
    \label{fig:anymod_optimization}
\end{figure}

The analyses above provide insights into Climate Change impacts on generation potential and demand but fall short of grasping the full complexity of energy systems that are spatially, temporally, and technologically interconnected. 
We therefore use the full C2E output in a stochastic energy system optimization with 20 historical or 20 future years, complemented with a deterministic optimization for all years separately (Fig. \ref{fig:anymod_optimization}).

We find that climate change following the SSP3-7.0 scenario reduces the need for all types of generation and storage, according to the stochastic optimization. 
This result is driven by the large demand reductions discussed above, and the choice to consider a fully-electrified heating sector in both periods. 
While the cost-optimal layout generates 6622 TWh of electricity on average during the historical period, only 5751 TWh are generated in the future.
The expected energy not served is negligible in both periods (0.0043\% historical; 0.0035\% future).

The system contains large storage capacities exceeding 10\% of annual demand, mostly in the form of hydrogen storage, that mitigate generation fluctuations. 
Generation is 73\% supplied by wind, 19\% by PV, and 6\% by dispatchable plants. 
Dispatchable plants operate when renewable generation is low, but demand is high, especially in winter. 
They are fueled with hydrogen that is produced when demand is low, predominantly during summer. 
Hydrogen storage thus serves as seasonal storage and over-compensates the identified changes in the seasonal cycle and inter-country balancing potentials. 
Batteries operate as short-term storage because their energy-related capacity costs are much higher. 
As a consequence, the average energy discharged from the hydrogen storage is about 90\% of its energy capacity while batteries discharge 2'900 times their energy capacity per year.
%In the future period, the basic mechanisms and aggregated generation shares of different technologies remain the same, but the average generation per year decreases to  5'751 TWh induced by the previously described drop in demand. 

The  changes in seasonality and spatial structure reported above do not relevantly impact the composition of renewable supply because (a) battery and hydrogen storage add substantial flexibility to balance out these changes, and (b) factors like installable capacity limits, and weather-independent demand remain unchanged and probably outweigh the more subtle changes in capacity factors.
Nevertheless, it remains an open question to what extent actual future energy systems can be operated as optimal as the model assumes. 
There is a risk that the combination of very large storage sizes and perfect foresight causes unrealistic power system operations by leveraging information that will never be available in the real world.

Running the same model deterministically for single years instead (blue shades in Fig. \ref{fig:anymod_optimization}) reveals substantial differences between years, though none of the single-year setups resembles the stochastic solution.
It follows that single-year analyses cannot approximate the ideal stochastic solution, potentially resulting in unreliable systems.
Finally, the convex hull overestimates capacities substantially relative to the stochastic optimization in both periods. 
Nevertheless, it otherwise confirms the general finding that climate change according to the SSP3-7.0 scenario would reduce capacity expansion needs. %Note that the convex hull is outside of the layered shapes because each axis in the radar plot aggregates multiple capacity variables.\footnote{To demonstrate why the most extreme points of the layered shapes do not define the convex hull, consider an example with two regions, A and B. For some technology, in the first result, region A has 3 GW, and region B has 0 GW; in the second result, region A has 0 GW, and region B has 3 GW. For both results and shapes, the plotted capacity value is 3 GW. Still, the convex hull takes the maximum across all results for each region separately, so 3 GW for each region, resulting in 6 GW total. Only having 3 GW in region A and 0 in region B could lead to loss-of-load in the second case; vice versa, only having 3 GW in region B in the first case.} 

\section{Conclusions}\label{sec:Conclusions}

We presented the Climate2Energy (C2E) framework to consistently assess energy sector impacts of climate variability and climate change. 
%C2E consistently formalizes the entire process from climate model output to energy system model input.
%It is openly available and can thus be used, critiqued, and modified. 
Combining C2E with dedicated hourly CESM2 climate simulations, we show that climate change impacts energy systems, most importantly via strong changes in demand for cooling ($>100$\%) and heating ($-10$\% to $-50$\%), as well as substantially lower hydropower generation potential in Southern Europe ($-10$\% to $-40$\%). 
Moreover, we document changes in seasonal cycles and spatial complementarity of all types of renewable generation that change the effectiveness of international transmission and long-term storage. 
Based on stochastic energy system optimization, however, we find that the total effect of these changes is reduced capacity expansion, highlighting the need to consider the complexity of energy systems that are spatially, temporally, and technologically interconnected.

%How this approach is better than existing ones
C2E leverages state-of-the-art conversion tools to improve the representation of different technologies in climate impact studies.
For instance, we use 3D wind and air density fields, thereby avoiding error-prone vertical extrapolation in the wind energy conversion. 
Moreover, the demand calculation utilizes country-specific building stock and temperature preference data and accounts for thermal inertia. 
The PV conversion includes panel orientation and temperature-dependent efficiency and the hydropower conversion is based on routed river discharge, as opposed to runoff, avoiding the need for simplified routing in the postprocessing. 
C2E also performs bias correction and its outputs are tailored to energy system model. 
In sum, C2E narrows the gap between climate modeling and energy system modeling.  

Nevertheless, we are aware of limitations. 
Owing to their coarse resolution, CMIP6-type global climate models like CESM2 can not capture important features like rivers or glaciers properly, and our results draw from one model, prohibiting evaluation of climate model uncertainty. 
We are restricted to one model because some conversion inputs are not available from existing databases like CMIP and CORDEX, requiring us to run CESM2 with dedicated outputs.
New versions of CMIP and CORDEX or other dedicated intercomparisons could solve this issue. 
Moreover, different choices could have been made with respect to covered sectors, technical detail or bias-correction.
Electric mobility, for instance, will have a sizeable weather-dependent energy footprint and wind park wakes may limit wind energy potentials.
Finally, the used stochastic energy system model still has two key limitations.
First, it strictly focuses on the power sector, omitting challenges and opportunities in other sectors.
Second, it assumes perfect weather foresight and might severely underestimate climate risks by planning for extreme events months in advance while skillful weather prediction at this timescale is unrealistic. 

While the limitations above matter, C2E contributes to overcoming them because it is open and flexible. 
It can, for instance, be used with higher resolution climate model output once available. 
Similarly, more sectors can be added as their relevance for the energy system grows, and different bias corrections, including multivariate ones \cite{vrac_multivariateintervariable_2015}, can be tested. 
While  we showed country-level results in Europe, output aggregation is flexible and the approaches are generally applicable globally.
We thus conclude that C2E enables us and others to better include climate change in energy system design, thereby narrowing the gap between climate and energy system modeling \citep{craig_overcoming_2022}. 

% Uncomment to get word count right
%\end{document}

\backmatter

\bmhead{Supplementary information}

\bmhead{Acknowledgements}
All authors are part of the ETH Zurich SPEED2ZERO initiative which received
support from the ETH-Board under the Joint Initiatives scheme. 
JW acknowledges support through the UiO-Equinor research collaboration.

\bmhead{Data and code (will become public when the paper is accepted)}

We provide C2E country-level output files for all realizations and gridded capacity factor fields for three realizations on zenodo (\url{https://doi.org/10.5281/zenodo.15269455}).
Hourly 3D climate model output files are at the order of TBs and can be provided upon reasonable request.
The anymod model and its input are available on zenodo (\href{https://doi.org/10.5281/zenodo.15276465}{10.5281/zenodo.15276466}).
The code is maintained on github and is openly available (\url{https://github.com/jwohland/CESM2energy}).

\bmhead{Author contributions}
J.W., L.G., L-B-W. and J.S. contributed to the conceptualization of the study. 
J.W., L.B-W and L.G. led the writing of the initial draft with feedback and improvements from all authors. 
U.B. performed the CESM2 simulations. 
J.W. implemented  the wind, solar and load conversion in C2E and created the corresponding figures. 
L.B-W implemented and evaluated the bias correction.
L.B-W. and F.DM. prepared the hydropower conversion model with guidance from J.W.. 
L.G. developed and ran the energy system model. 
R.K. acquired funding for the study. 

\clearpage
\bibliography{climate2energy}

\clearpage
\begin{appendices}

\setcounter{page}{1}
\setcounter{figure}{0}
%\resetlinenumber
\renewcommand{\thetable}{S\arabic{table}}%
\renewcommand{\thefigure}{S\arabic{figure}}%

\textbf{\LARGE Supplementary Material\newline}
    
\vspace{2cm}
    
\begin{center}
    \textbf{Climate2Energy: a framework to consistently include climate change into energy system modeling}
\end{center}

\clearpage
\section{Detailed hydropower documentation} \label{sec:hydro_details}

In this section, we detail each step of the hydropower conversion tool developed within the scope of Climate2Energy, including pre-processing, conversion, and evaluation. 

\begin{figure}[h]
    \centering
    \includegraphics[width=\linewidth]{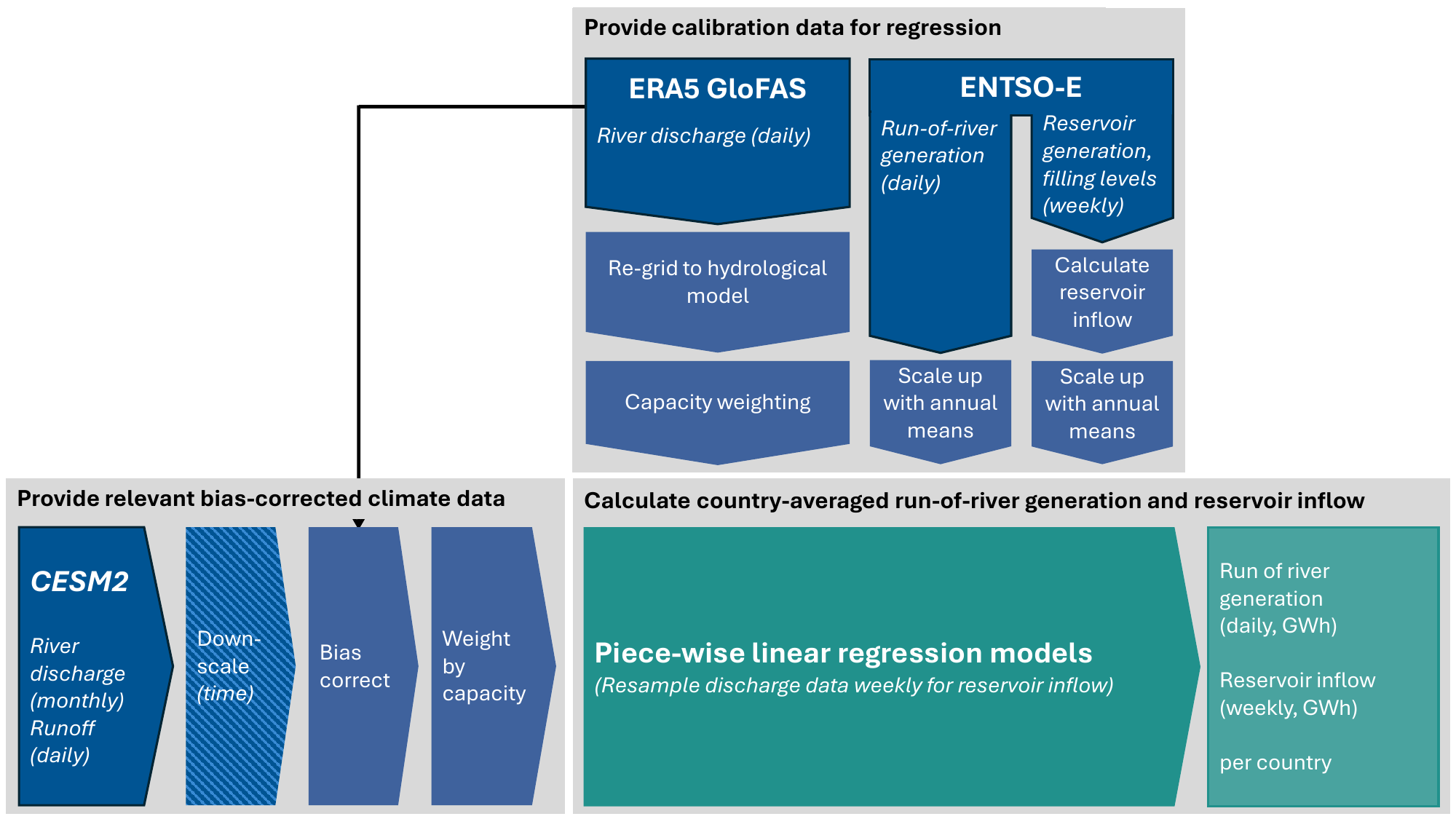}
    \caption{Illustrative schematic of the hydropower conversion tool. Key steps are separated by color: provision of relevant climate and calibration data (blue), and computation and post-processing of hydropower energy in generation and inflow (green). Temporal down-scaling is shown in hashed blue, since it is necessary only due to constraints in climate model output. This conversion tool provides CSV files on country granularity, in line with the general Climate2Energy pipeline.}
    \label{fig:hydro_pipeline}
\end{figure}

Figure \ref{fig:hydro_pipeline} provides an overview of the full conversion process: First, we pre-process the climate output of CESM2, namely daily river runoff and monthly river discharge. 
This pre-processing consists of temporal downscaling, bias correction, and capacity weighting. 
Then, the calibration data is preprocessed: First, the historical river discharge from ERA5 is re-gridded to the hydrological model of CESM2 and capacity-weighted \cite{harrigan_glofas-era5_2020}. 
Next the reported historical run-of-river hydropower generation is retrieved from ENTSO-E for the same period \cite{hirth_entso-e_2018}. 
Additionally, reservoir inflows are calculated from filling levels and reservoir generation data from ENTSO-E. 
All pre-processing is detailed in the following subsection \ref{preproc}. 
Finally, hydropower energy from CESM2 is calculated using a piece-wise linear regression fitted to the calibration data (see Section \ref{convert}). 
Since ENTSO-e reports daily values for run-of-river and weekly values for reservoirs, the temporal resolution of the ERA5 and CESM2 is resampled accordingly. 

\begin{figure}[h]
    \centering
    \includegraphics[width=\linewidth]{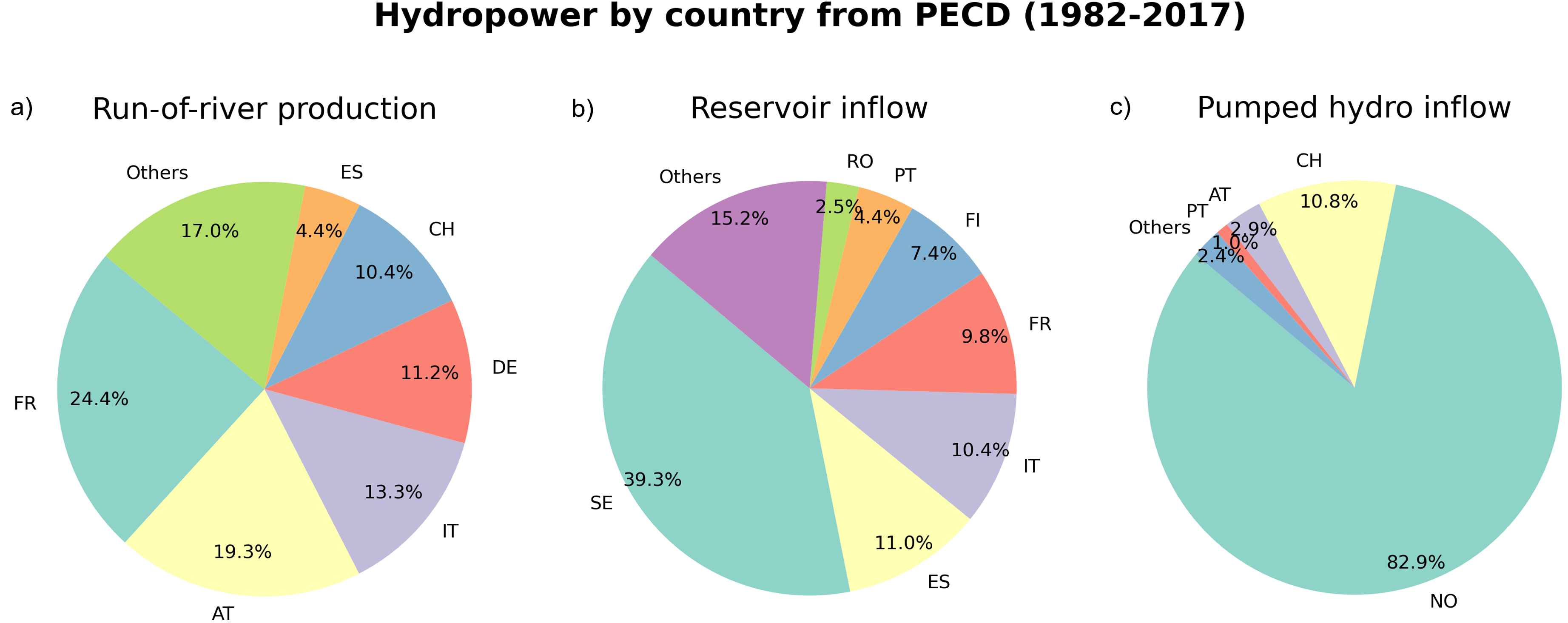}
    \caption{Production share of three hydropower technologies: a) run-of-river generation, b) hydro reservoir inflow, and c) pumped hydro inflow. For run-of-river, "Others" account for the other 28 countries reported in the PECD Hydropower database. Energy values for each country are summed over the full span of the PECD Hydropower database, i.e., 1982-2017, and using the output based on 2025 capacities.}
    \label{fig:PECD_hydro_share}
\end{figure}

The described process is performed separately for each country. 
For the sake of simplicity, we only consider countries with noteworthy hydropower production, as reported from PECD \cite{de_felice_entso-e_2022}. 
The production share of the different hydropower technologies in the considered countries is shown in \ref{fig:PECD_hydro_share}. 
For run-of-river, we include Austria, France, Germany, Italy, Spain, and Switzerland, covering 83\% of European production. 
For reservoirs, we include Austria, Bulgaria, France, Italy, Norway, Portugal, Romania, Spain, Sweden and Switzerland, covering 80\% of reservoir inflows and 99\% of pumped hydro inflows. 
Despite a considerable share of reservoir inflows, we excluded Finland due to a lack of corresponding generation data on the ENTSO-E Transparency Platform.

\subsection{Preprocessing the datasets} \label{preproc}

\subsubsection{Downscaling monthly CESM2 discharge to daily values}

 Runoff is currently the most commonly used climate model variable to estimate hydropower generation \cite{van_der_most_temporally_2024,andresen_validation_2015}. 
 However, runoff must be routed through a hydrological model to capture water flows in a catchment area since it only includes local precipitation and snow melt, but omits inflows from adjacent grid cells. 
 Here, we base our hydropower modeling on river discharge, a climate model variable which is based on internally routed runoff within the climate model itself, by using the Model for Scale Adaptive River Transport (MOSART,\cite{li_physically_2013}). 
 This physically based model routes surface runoff across hill slopes and discharges it, along with subsurface runoff, while ensuring spatial links between all model parameters. Additionally, it has the benefit of a 0.5 $\cdot$ 0.5$^\circ$ resolution, almost 4 times higher than the atmospheric grid of CESM2.1.2 \cite{danabasoglu_community_2020}.

We only archived river discharge at a monthly temporal resolution when running the climate model because we were unaware that we would need it for the hydropower conversion when performing the climate simulations. 
Since the hydropower conversion model needs daily values, we temporally down-scale it using daily runoff values. 
To perform the down-scaling, we assume that the daily river discharge evolution within a month follows that of runoff, keeping the monthly river discharge means unchanged. 
Specifically, we multiply the monthly river discharge values $Q'(\Vec{x},\text{month})$ with the daily runoff $R(\text{month}, \text{day})$ and scale with the monthly mean runoff $R_\text{mean}(\text{month})$ :

\begin{equation}
    Q(\Vec{x},\text{month}, \text{day}) = Q'(\Vec{x},\text{month}) \cdot \frac{R(\text{month}, \text{day})}{R_\text{mean}(\text{month})}, 
\end{equation}

where $Q(\Vec{x},\text{month}, \text{day})$ is the derived daily discharge and $\Vec{x}$ denotes the spatial location. 
This downscaling step can easily be avoided in future versions of the dataset by including higher resolution river discharge in the output list. 

\subsubsection{Regridding ERA5 to fit CESM2 grid and bias correction}

Since the grid resolution of ERA5 is ten times higher than the resolution of CESM2 river discharge, we must re-grid the data to perform the bias correction for two reasons: 

\begin{enumerate}
\item To keep consistency with the other C2E components, since bias correction was also performed before aggregation for the other C2E variables. If we want to perform grid-by-grid bias correction, the two grids therefore need to be equal.
\item To compute the capacity weighting (see section \ref{cap}) consistently across datasets. With a lower resolution, weights are less precise, which might lead to an over- or underestimation of the discharge. Therefore, without re-gridding ERA5 to CESM2, the latter could be biased relative to the former.
\end{enumerate}

The re-gridding is performed by coarsening the ERA5 grid by a factor of 10 in both latitude and longitude directions, so that it matches the CESM2 grid. 
After re-gridding, CESM2 is bias corrected with ERA5 using quantile mapping as seen in Section \ref{sec:Methods} of the main paper. 
We use the same historical time range as for the other conversions, namely 1995-2014.

\subsubsection{Calculating reservoir inflow from filling levels and generation}

Since hydropower reservoirs have storage capacity, energy generation and river discharge are decoupled in time, which makes energy inflows (in GWh) a more meaningful output. 
However, this value is not available directly for ENTSO-e, and we therefore calculate it with a modified version of \cite{gotske_future_2021}:

\begin{equation}
    V_{w+1}=V_w + E_w-\sum_{h=1}^{168}G_h {\frac{1}{\eta}}
    \label{eq:filling_level}
\end{equation}

where $V_w$ is the reservoir filling level at week $w$, $V_{w+1}$ the same at week $w+ 1$, $E_w$ is the energy inflow, $G_h$ is the electricity generation at hour h, and $\eta$ is the conversion efficiency of the hydropower plant. 
The original source additionally considered an energy loss term. 
We follow a different approach by accounting for energy losses in the efficiency term, and the efficiency value considered is 94.9\%. 
The weekly energy inflow $E_w$ is therefore approximated by this equation from hourly electricity generation and weekly reservoir filling level at ENTSO-E.

\begin{figure}[h]
    \centering
    \includegraphics[width=\linewidth]{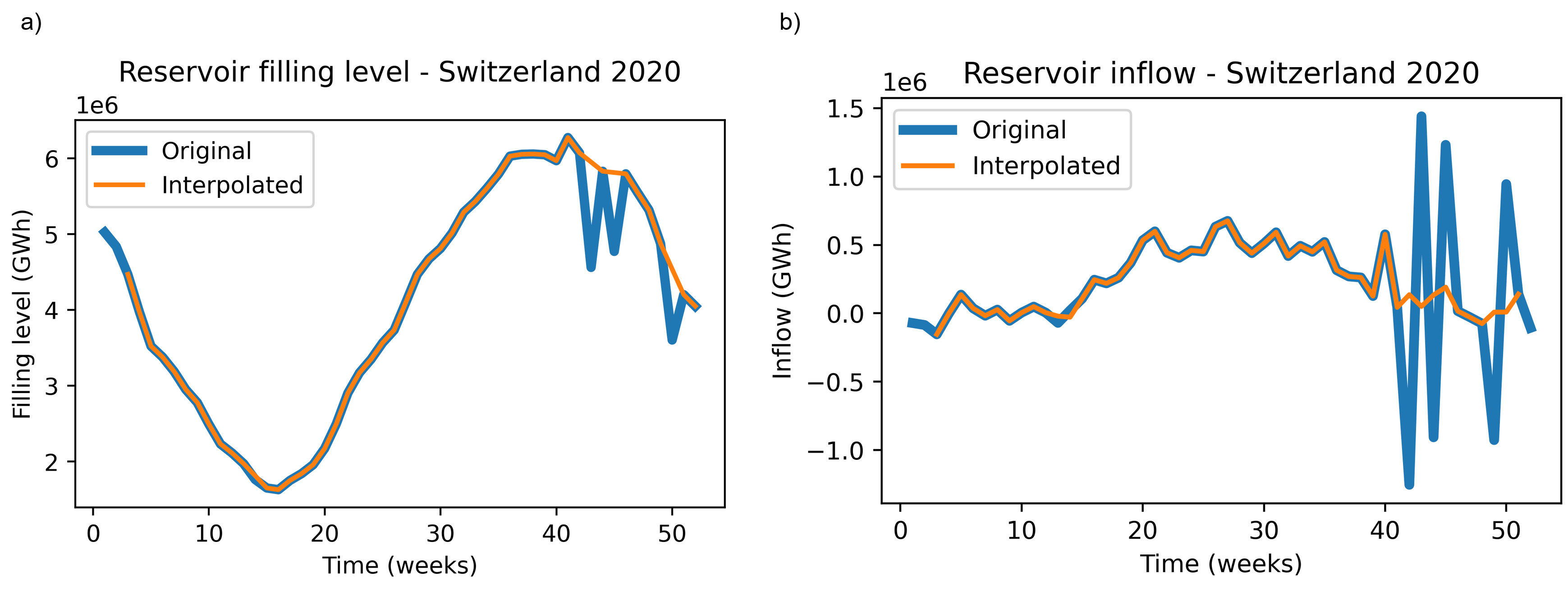}
    \caption{Historical reservoir inflows in panel b) are calculated with Equation \ref{eq:filling_level}, that uses as input the historical reservoir filling level, as reported from ENTSO-E. a) shows an example for Switzerland in 2020. Some reporting issues between weeks 40 and 50 in the filling level resulted in unrealistic inflow spikes, including negative inflows. We replace these outlier values in the filling levels with an interpolated value. As a result, the inflow coming from the "Interpolated" values does not oscillate as in the "Original".}
    \label{fig:interp}
\end{figure}

Certain countries show unrealistic filling level fluctuations at certain time steps $n$ that lead to negative energy inflow values (see Fig. \ref{fig:interp}). 
This artifact is likely due to reporting errors in the ENTSO-e datasets of these countries and we resolve it by replacing erroneous values (time step $n$) with smoothed interpolations between values before and after the respective time step (i.e., at step $n-1$ and $n+1$).

\subsubsection{Scaling up}

Because of challenges in the reporting of actual generation, ENTSO-e transparency time series data do not always have the right absolute values. 
However, the ENTSO-e power statistics provide more reliable monthly data, but are only available during the last three years. 
Assuming that ENTSO-e transparency data, even with wrong absolute values, is internally consistent in terms of the temporal dynamics of the time series, we scale up each time series with the annual mean generation over the last three years per country.

\subsubsection{Capacity weighting}\label{cap}

We compute a weighted mean river discharge at the country level $Q_c(t)$, where the weights correspond to the fraction of installed hydropower capacity per grid cell for a given country. 
This is done by calculating a standardized map of hydropower plant locations, weighted by hydropower capacity, which is then summed over each country:

\begin{equation}
    Q_{c}(t) = \sum_{x \in c}\frac{Q_i(\Vec{x}, t)C'_c(\Vec{x})}{C_c}
\end{equation}

where $t$ is time, $Q_i(\Vec{x}, t)$  is the discharge at location, $\Vec{x}$, $C'_{c}(\Vec{x})$  is the weight of grid cell i and $C_{c}$ is the total capacity of all grid cells in country $c$ \citep{rozsai_jrc-idees-2021_2024}.

\subsection{Conversion model} \label{convert}

\subsubsection{Piece-wise linear regression}

Hydropower generation $P$, energy $E$, and river discharge $Q$ are, through first principles, linked in the following way:

\begin{equation}
    P = \frac{\mathrm{d}E}{\mathrm{dt}} = \frac{\mathrm{d}m}{\mathrm{dt}} g h =\rho Qgh \propto Q
\end{equation}

where $m$ is the mass, $g$ is the gravitational constant, $\rho$ is water density, and $h$ is the height. We can thus model $P$ as a linear function of $Q$ for a hypothetical hydropower plant with unlimited capacity and perfect efficiency. 
Note that $P$ can either be in the form of reservoir inflow or run-of-river generation. 

To improve the accuracy of the linear function, we add three constraints. 
First, we perform a piece-wise linear regression, where values above the 75th percentile are fitted with a linear intercept to account for spilling at extreme discharge levels. 
Second, we allow for a linear intercept $b_1$, forced to be negative or 0, for values under the 75th percentile. 
This is because some countries show no generation when there is some discharge. 
There are several possible explanations for such an observation. 
One reason could be the relatively coarse spatial resolution of the discharge data. 
For instance, if most of the discharge occurs in areas of a grid cell without installed capacity, the observation might show limited or no power generation. 
Another plausible explanation is the existence of a discharge threshold that must be exceeded before power can be generated.
Finally, since negative hydropower energy is not physical, we force all values of $E$ to be zero if the linear regression would otherwise give a negative result.
This considerations yield the following equation:

\begin{equation}
P(Q) = 
\begin{cases}
    0, & \text{if } Q < \frac{b_1}{a_1} \\
    a_1 Q + b_1, & \text{if } \frac{b_1}{a_1} \leq Q < Q_{75^\mathrm{th}} \\
    a_2 Q + b_2, & \text{if } Q \geq Q_{75^\mathrm{th}}
\end{cases}
\end{equation}\label{eq:lin_trans}

\begin{align*}
&\text{Subject to:} \\
&a_1 Q_{75^\mathrm{th}} + b_1 = a_2 Q_{75^\mathrm{th}} + b_2 \qquad \text{(continuity condition)} \\
&\text{with } a_1 > 0,\ a_2 > 0,\ b_1 \leq 0
\end{align*}

We determine the parameters above using run-of-river and reservoir data from ENTSO-e as described above, and discharge from ERA5 for the period 2017-2022 as calibration data. 
The parameters are calibrated separately for each hydropower technology considered, since the ENTSO-E data is different, and the hydropower locations in the capacity weighting would lead to different river discharge time series.

\begin{figure}[h]
    \centering
    \includegraphics[width=\linewidth]{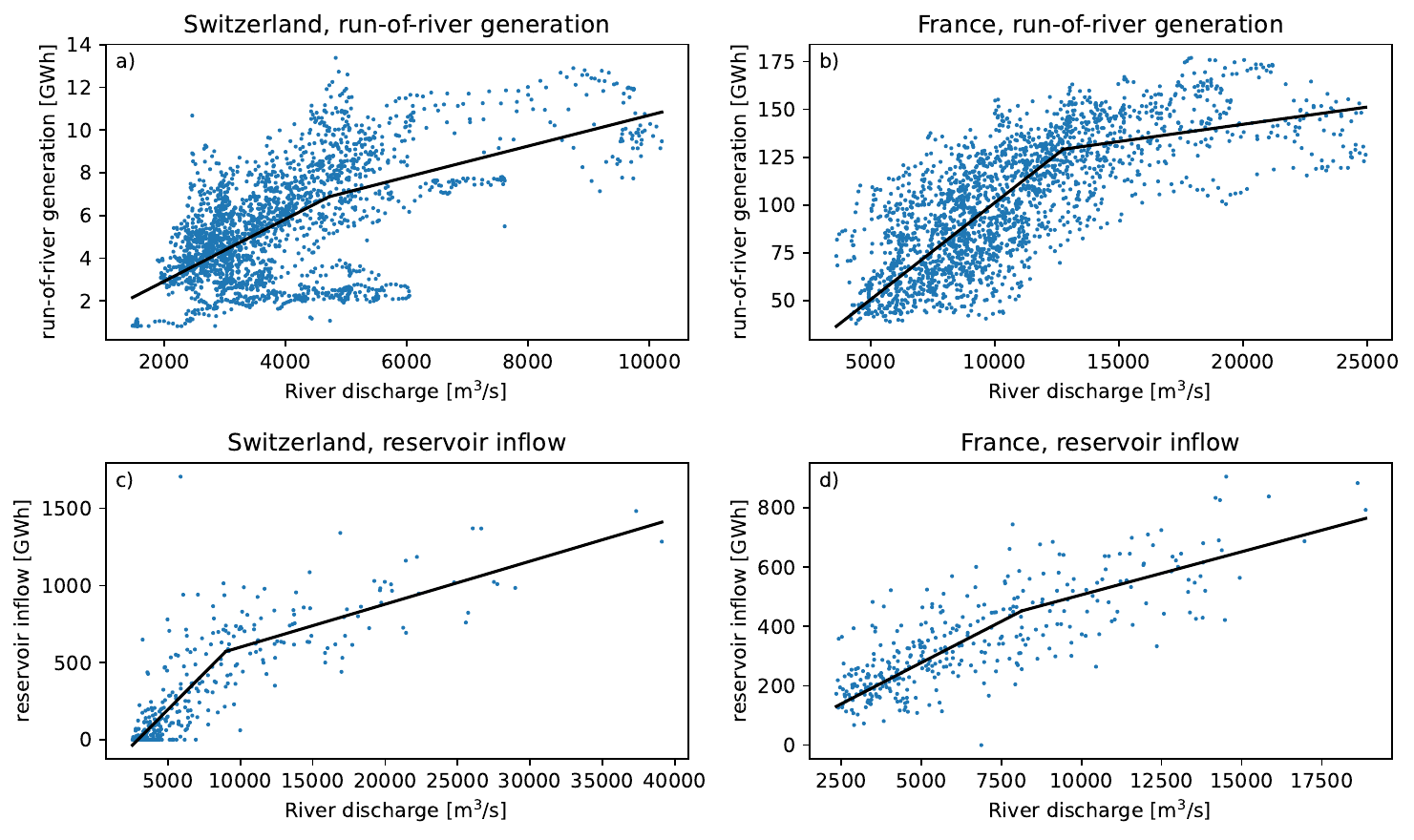}
    \caption{Linear regression data and fitted piece-wise transfer function for (a,c) Switzerland and (b,d) France, for (a,b) run-of-river generation and (c,d) reservoir inflow.}
    \label{fig:reg_lin}
\end{figure}

Figure \ref{fig:reg_lin} shows the training ENTSO-e and ERA5 data used to determine the piecewise linear regression, along with the regression itself, for two selected countries: Switzerland and France.
It shows that the piecewise linear model captures the main characteristics of the relationship between river discharge and generation (or inflow) and is therefore useful to understand large-scale changes and trends. 
Nevertheless, the data exhibits considerable scatter, highlighting that other factors also play a role. 
We will therefore evaluate the hydropower conversion in terms of its ability to capture different characteristics of generation and inflow in the following.

\subsection{Evaluating the hydropower conversion}

Due to the novel nature of this conversion tool, it is necessary to evaluate its ability to robustly convert river discharge to hydropower energy. Here, we first compare the final hydropower energy of ENTSO-E, ERA5 and CESM2 for both reservoirs and run-of-river, before evaluating the regression model and the differences between ERA5 and CESM2 river discharge.

\subsubsection{Comparing ERA5 and CESM2 based hydropower estimates to reported ENTSO-E data}

\begin{figure}[h]
    \centering
    \includegraphics[width=\linewidth]{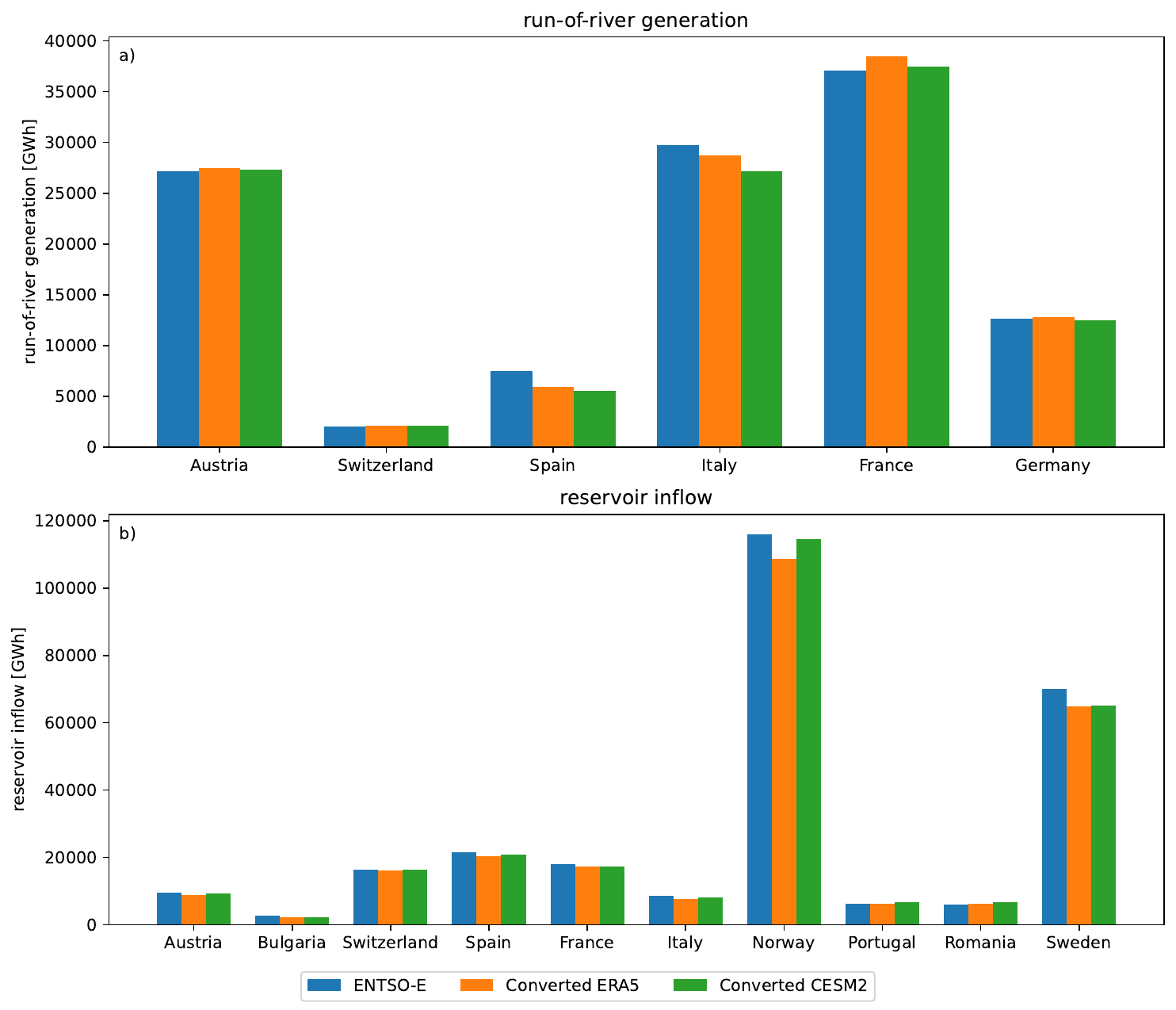}
    \caption{Annual mean (a) run-of-river generation and (b) reservoir inflow in GWh for ENTSO-E (2017-2022), ERA5 (1995-2022) and CESM2 (1995-2014) for all the countries considered in the hydropower conversion. }
    \label{fig:an_prod}
\end{figure}

We begin with a comparison of annual means. 
Figure \ref{fig:an_prod} shows the annual hydropower energy generation for run-of-river and reservoirs for each country. 
Here, we find an overall good match between the ENTSO-E data used for calibration and the ERA5 and CESM2 data converted through the piece-wise linear regression. 
The mean relative error between the two converted datasets and ENTSO-E is 6.0\% for run-of-river generation and 6.2\% for reservoirs, with values spanning 0.3\% and 26\%.

Note that the considered periods are not identical because of limited data ENTSO-E data availabilty and choices in the CMIP climate scenario definitions. 
While we use ERA5 and ENTSO-E during 2017-2022 to determine the parameters of Eq. \ref{eq:lin_trans}, we apply those parameters to the historical CESM2 simulation (1995-2014) and compare the results to the corresponding period in ERA5. 
In an ideal setup, all data sources would cover the same period. 
Nevertheless, the approach chosen here is plausible because CESM2 is a non-initialized climate model, implying that year-to-year variability in the model is not synchronized with the real world.

\begin{figure}[ht!]
    \centering
    \includegraphics[width=0.9\linewidth]{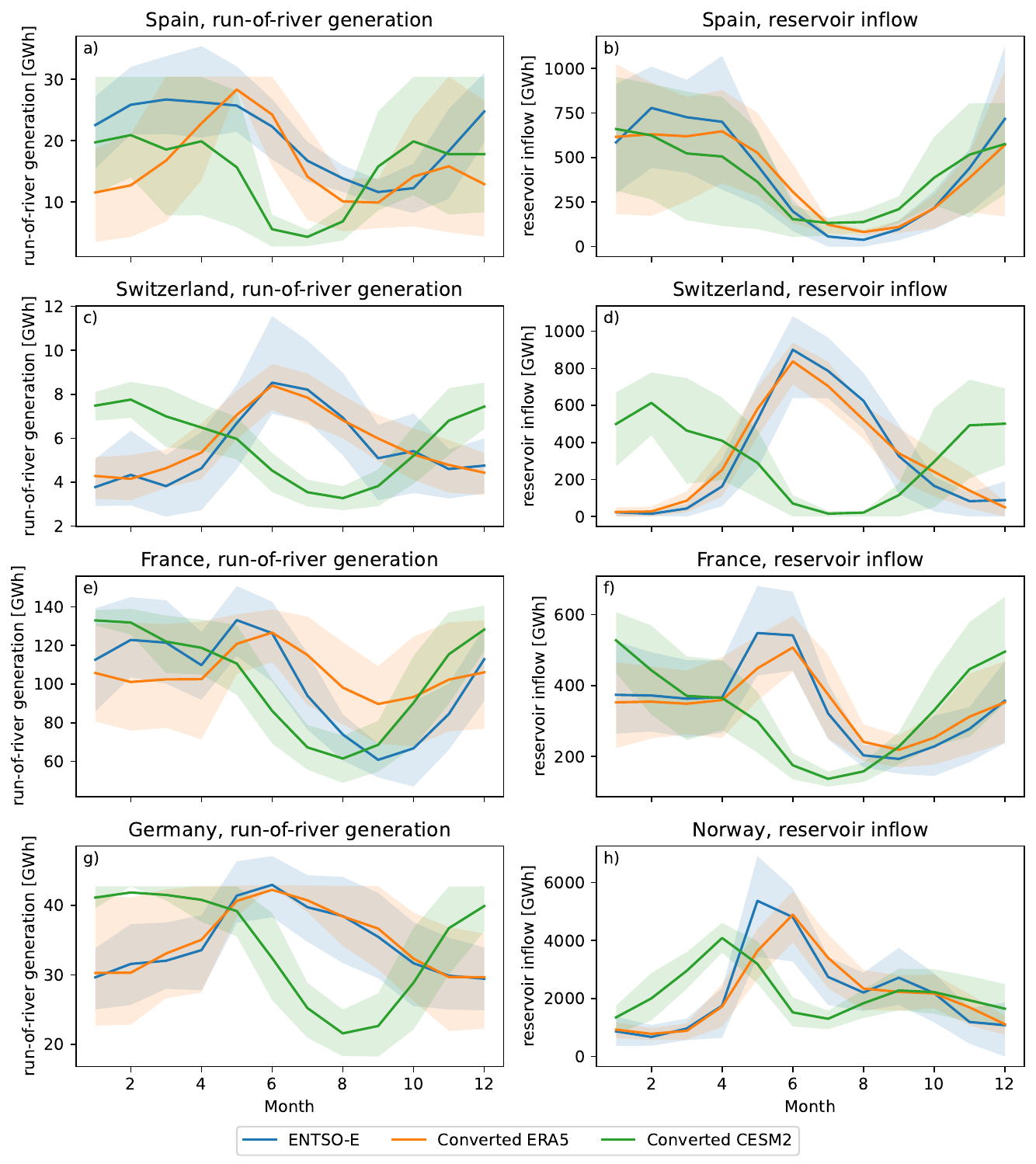}
    \caption{Monthly mean (a,c,e,g) run-of-river generation and (b,d,f,h) reservoir inflow in GWh for (a,b) Spain, (c,d) Switzerland, (e,f) France, (g) Germany and (h) Norway for ENTSO-E (2017-2022), ERA5 (1995-2022) and CESM2 (1995-2014). Shaded areas correspond to the 20$^{\text{th}}$ and 80$^{\text{th}}$ percentile values. }
    \label{fig:season_cycle}
\end{figure}

We proceed with a comparison of the seasonal cycles (Fig. \ref{fig:season_cycle}). 
Since the linear regression omits seasonality when performing the conversion, the seasonal cycle of hydropower energy is an independent test. 
However, caution must be taken when interpreting these results, due to the short sample of ENTSO-E data (5 years). 

Nevertheless, the observed ENTSO-E generation and the ERA5-based modeled generation exhibit similar seasonal patterns, with consistently overlapping uncertainty bands.
This similarity suggests that the piece-wise linear model is capable of reproducing the measured seasonal evolution when provided with high-quality input data.
In contrast, the CESM2-based modeled generation features different seasonal patterns. 
In general, the seasonal cycle of reservoir inflows and run-of-river generation in CESM2 is similar across countries, with a decrease in summer compared to winter. 
This suggests precipitation-driven hydropower energy, rather than the snowmelt-driven energy expected in mountainous countries like Switzerland and Norway. 
The close agreement between the observed data and the ERA5-based model indicates that the piecewise linear hydropower conversion model effectively captures seasonal hydrological variations and translates them into seasonal energy production. 
Therefore, the discrepancy seen in the CESM2-based generation is not due to the hydropower model itself, but rather reflects limitations in the CESM2 climate model’s ability to represent seasonal hydrological patterns accurately.

We also observe variations between and even within countries: 
while all three data sets agree on the seasonal cycle in Spain for reservoirs (Figure \ref{fig:season_cycle}b), they follow different cycles for Spanish run-of-river (Figure \ref{fig:season_cycle}a). 
Nevertheless, ENTSO-E and CESM2 share some essential characteristics like lower levels in summer and higher levels in winter. 
This is also the case for run-of-river in France (Figure \ref{fig:season_cycle}e), where ENTSO-E and CESM2 seasonally correspond well, but with a time lag of roughly a month and overlapping uncertainty bands. 
In Switzerland (Figures \ref{fig:season_cycle}c,d), however, ERA5 and ENTSO-E are consistent, with CESM2 following an almost inverse cycle. 
A similar discrepancy can be seen in run-of-river for Germany (Figure \ref{fig:season_cycle}g). 
Those results likely indicate the limited ability of the climate model to capture precipitation, snow accumulation and snow melt in the alps properly while they could also be impacted (to a lesser degree) by underlying trends and the short sample of ENTSO-E observations that include particularly dry periods. 
Finally, reservoirs in France and Norway (Figures \ref{fig:season_cycle}f,h) have partly similar seasonal cycles in all three data sets, but CESM2 does not capture the increased inflow in late spring and summer that is present in both ENTSO-E and ERA5.

Overall, the seasonal assessment provides evidence that the chosen approach reproduces reported seasonal cycles well while also emphasizing climate model limitations. 
Thus, caution must be taken when interpreting CESM2-based hydropower on a time scale shorter than the annual one. 
While artificially separating calibration data by season would lead to an improvement in seasonal results, it would also break with the first principles governing our simple and explainable linear regression (since there is no physical reason for different translational coefficients between discharge and hydropower in different seasons). 
Additionally, since the model is agnostic to both the climate model and calibration data, better results are expected with higher resolution and more reliable climate model river discharge and longer time series of calibration data. 

\begin{figure}[ht]
    \centering
    \includegraphics[width=\linewidth]{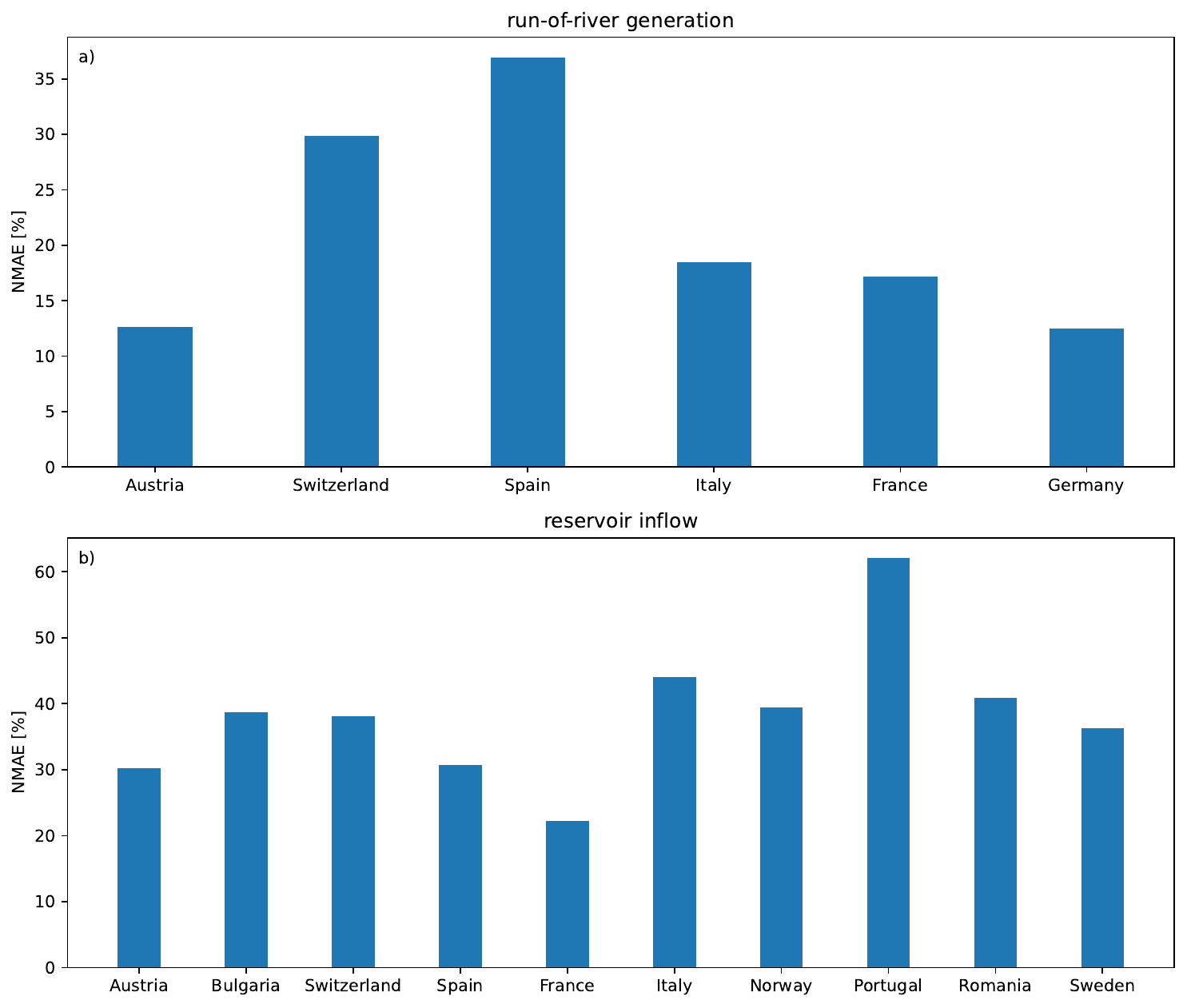}
    \caption{Normalized Mean Absolute Error between ERA5 converted hydropower and ENTSO-E in the time range 2017-2022 for (a) run-of-river generation and (b) reservoir inflow for all the countries considered in the hydropower conversion.}
    \label{fig:nmae}
\end{figure}

\begin{table}[h]
    \centering
    \begin{tabular}{c|c c }
       Country  &  Run-of-river generation & Reservoir inflow\\
       \hline
         Austria & 0.85 & 0.87 \\
         Bulgaria & & 0.79 \\
         Switzerland & 0.50 & 0.84\\
         Spain & 0.58 & 0.89 \\
         France & 0.75 & 0.78\\
         Germany & 0.75 & \\
         Italy & 0.80 & 0.71\\
         Norway & & 0.77 \\
        Portugal & & 0.75 \\
        Romania & & 0.55 \\
        Sweden & & 0.75 \\
         
    \end{tabular}
    \caption{Spearman correlation between ERA5 converted hydropower and ENTSO-E in the time range 2017-2022}
    \label{tab:corr}
\end{table}

As a last evaluation step, we assess performance on a per timestep basis (Fig. \ref{fig:nmae}).
We evaluate the piece-wise linear regression using ERA5 discharge relative to ENTSO-E during the period of overlap (2017-2022) in terms of  Normalized Mean Absolute Errors (NMAE) per timestep (daily for run-of-river and weekly for reservoirs). 
Note that the choice of metric implies that errors can not compensate each other, for example, when a generation peak is captured accurately but too late or too early.
Our results show that run-of-river NMAE lie between 12 and 37 \%, while reservoirs have NMAE between 22 and 62 \%, with Portugal being the only country above 50 \%. 
The magnitude of errors aligns with previous studies that modeled hydropower based on climate data, \cite{van_der_most_extreme_2022}, with RMSE ranging up to 18\% for run-of-river and 44\% for reservoirs. 
These errors, while non-negligible, indicate that the conversion tool gives results that, on average, are in the same magnitude as the ground truth at any given time step. 
Given that hydropower plays a more minor role in the European energy system (around 11\% of the EU energy mix in 2023 \cite{eurostat_gross_2022}), with limited potential for capacity expansion, this is deemed an acceptable level of error. 
However, caution will need to be taken if data is applied to single countries or when focusing on hydropower alone.

Additionally, the Spearman rank correlations between ERA5 and ENTSO-E exceed 0.7 in most cases (Table \ref{tab:corr}), confirming the general suitability of the approach. 
These correlations are comparable (within 0.1) or higher (increases by up to 0.63) than values reported in \cite{van_der_most_temporally_2024}.

To conclude, our approach provides good estimates on annual timescales with minor deviations between modeled and reported generation or inflow. 
Moreover, it reproduces the seasonal evolution of generation and inflows correctly when fed with high quality reanalysis inputs, yet inherits limitations from the climate model that imply a less realistic representation in some cases. 
Lastly, the approach performs reasonably well on high temporal resolution, as evidenced by comparatively high spearman rank correlations and elevated yet acceptable Normalized Mean Absolute Errors. 
We therefore argue that the approach is suitable for the assessment carried out in the main paper, while underlining the need for higher-resolution and more accurate climate inputs and refinements in the conversion methodology. 

\subsubsection{Comparing river discharge in ERA5 and CESM2}

\begin{figure}[h]
    \centering
    \includegraphics[width=\linewidth]{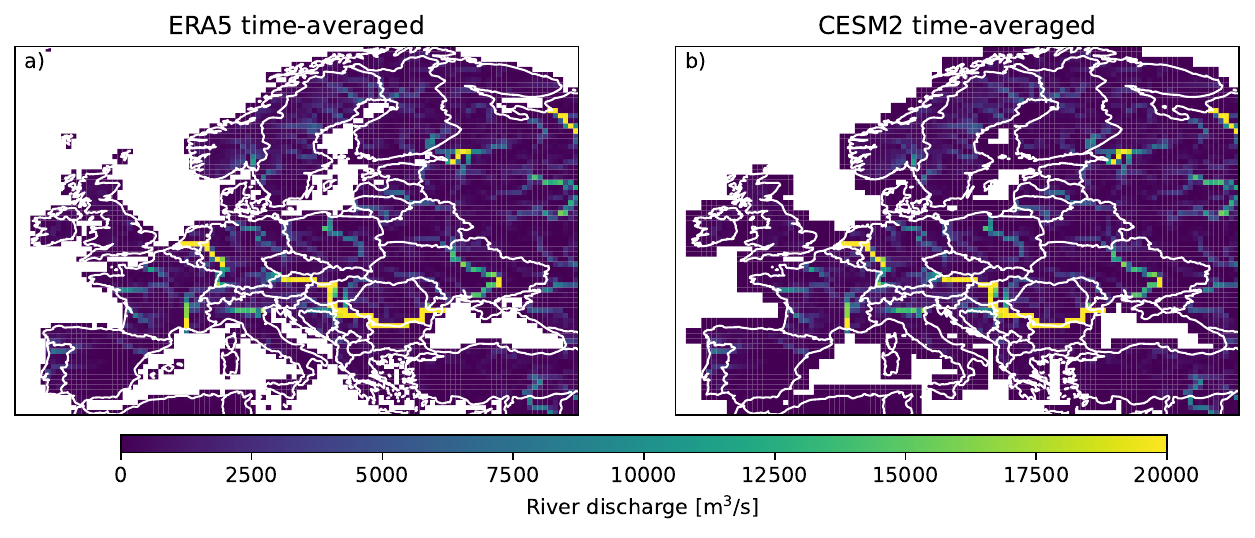}
    \caption{Maps of mean river discharge over Europe for ERA5 and CESM2 in the historical time period 1995-2014.}
    \label{fig:mean_dsch}
\end{figure}

To gain further insights into the climate-model limitations, we now step back from the hydropower conversion and compare  bias-corrected CESM  discharge to re-gridded ERA5 discharge during the historical period (1995-2014). 
Due to the quantile delta mapping bias correction being performed on each grid cell separately, any map of median and/or quantiles will, by definition, be equal between the two datasets. 
Nevertheless, time-averaged ERA5 and CESM2 are shown in Figure \ref{fig:mean_dsch} to highlight how the datasets capture the spatial distribution of discharge in the main European river networks. 

\begin{figure}[h]
    \centering
    \includegraphics[width=\linewidth]{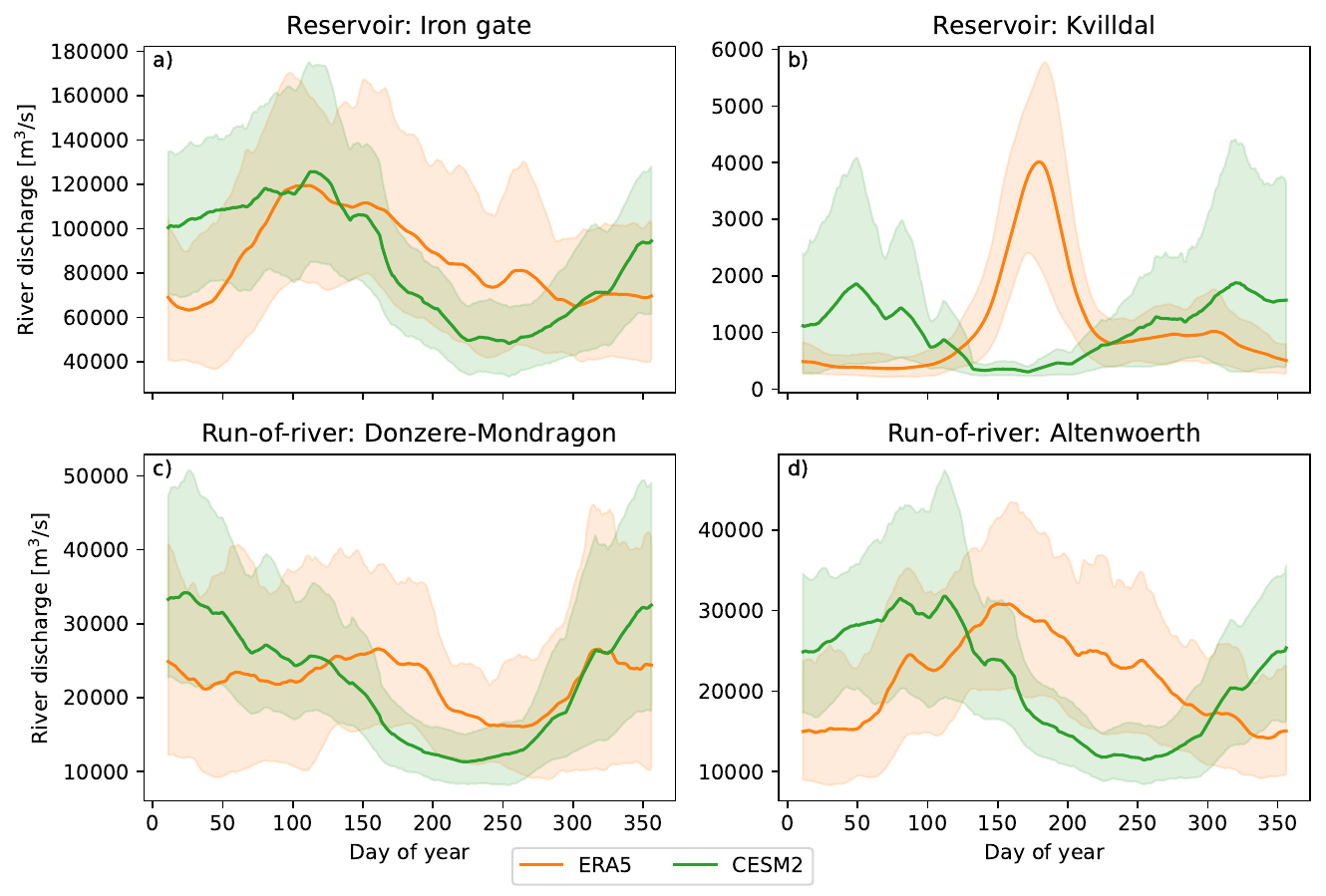}
    \caption{Seasonal cycle of river discharge in the two largest (a,b) reservoir power plants and (c,d) run-of-river power plants. Shaded areas correspond to the 10$^{\text{th}}$ and 90$^{\text{th}}$ percentile values.}
    \label{fig:season_dsch}
\end{figure}

Additionally, Figure \ref{fig:season_dsch} shows the seasonal cycle of the grid cell corresponding to the location of the two largest reservoirs and run-of-river power plants in Europe. In all stations, except Kvilldal, the seasonal cycle of river discharge seems to correspond partly between ERA5 and CESM2 data, at least within one standard deviation. However, in all four cases, CESM2 outputs a similar cycle, which again points to errors in modelling snowmelt-driven discharge peaking in spring-summer. This discrepancy is particularly striking in Kvilldal, a grid cell located at the west coast of Norway. Here, the challenge of capturing the complex topography of Norwegian fjords in a coarse resolution grid might also play a role. 

\section{Effect of including variable air density in wind energy conversion}\label{app:density}

We include evolving air density $\rho$ in the conversion to wind energy capacity factors using the method outlined in \citet{svenningsen_power_2010} that was inspired by the IEC 61400-12 norm and compare to using the IEC standard density $\rho_{IEC}=1.225 kg/m3$. 

While wind energy density
\begin{align}
    \text{WED} = \frac{1}{2} A \rho s^3
\end{align}

is linear in air density $\rho$, it does not make sense to scale capacity factors linearly with air density as doing so would also modify the rated capacity. 
Instead, we compute a pseudo wind speed $s'$ that captures the effects of air density changes, and use it with the turbine power curve to compute the capacity factor $CF(s')$.
To compute $s'$, we demand that
\begin{align}
    \text{WED}(\rho_\text{IEC}, s’) = \text{WED}(\rho, s)
\end{align}

meaning that the pseudo-wind speed speed combined with the standard air density $\rho_{IEC}$ leads to the same wind energy density as using the actual density $\rho$ with the actual wind speed. 
Rearranging terms yields

\begin{align}
    s’ = \left(\frac{\rho}{\rho_\text{IEC}}\right )^{1/3}  s.
    \label{eq:pseudo-wind}
\end{align}

\subsection{Mean capacity factors: relevant change in the climatology and weak but consistent change in change signal}

\begin{figure}[ht!]
    \centering
    \includegraphics[width=.8\textwidth]{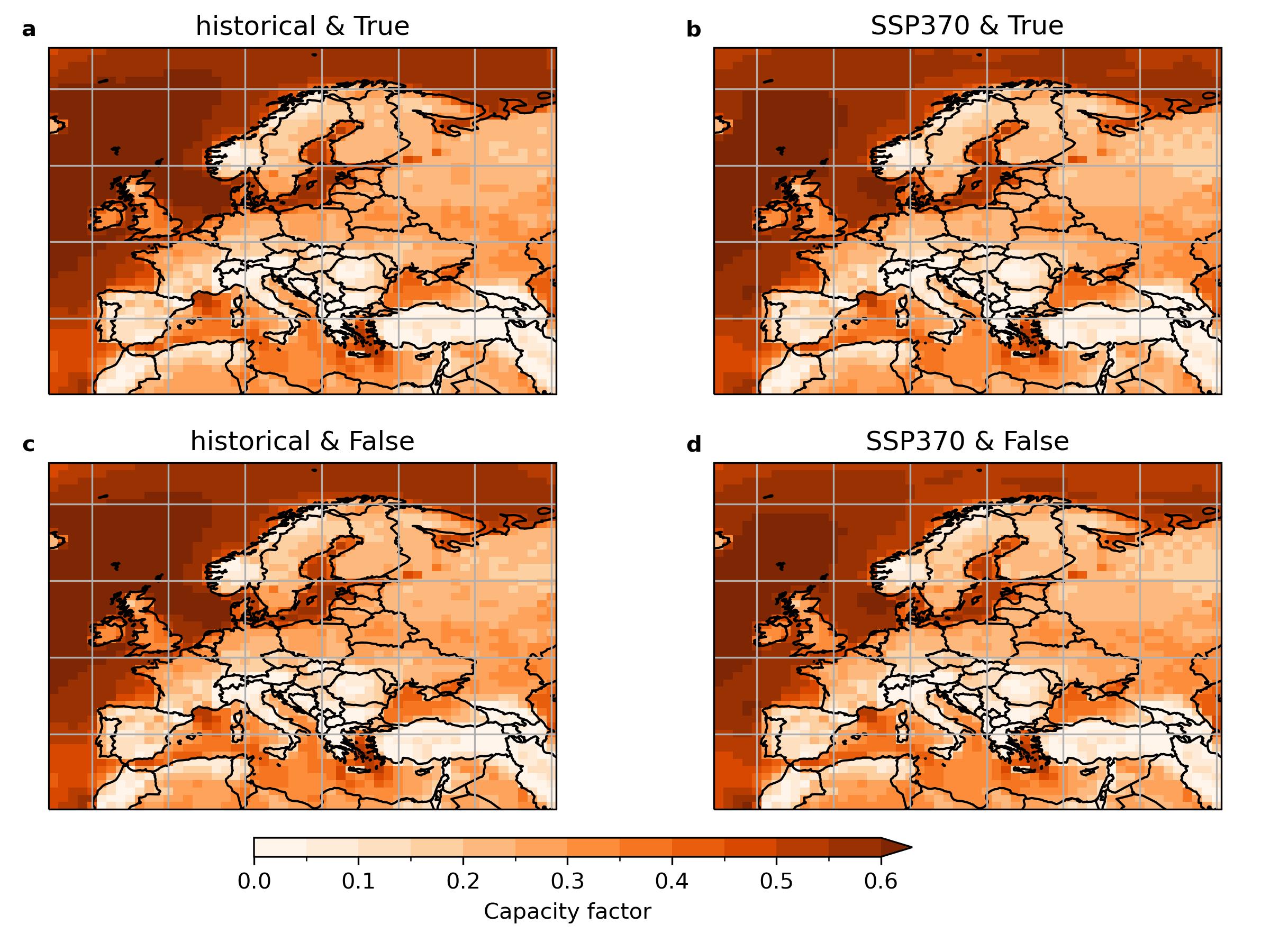}
    \caption{Mean capacity factors for the median SWT120\_3600 turbine under historical (a, c) and future (b, d) climate conditions with (a, b) and without (c, d) air density correction. Results are shown for realization A and bias corrected with realization A and averaged over the entire 20y period (1995-2015 historical; 2080-2100 SSP3-7.0).}
    \label{fig:app_mean_maps_density}
\end{figure}

\begin{figure}[ht!]
    \centering
    \includegraphics[width=\textwidth]{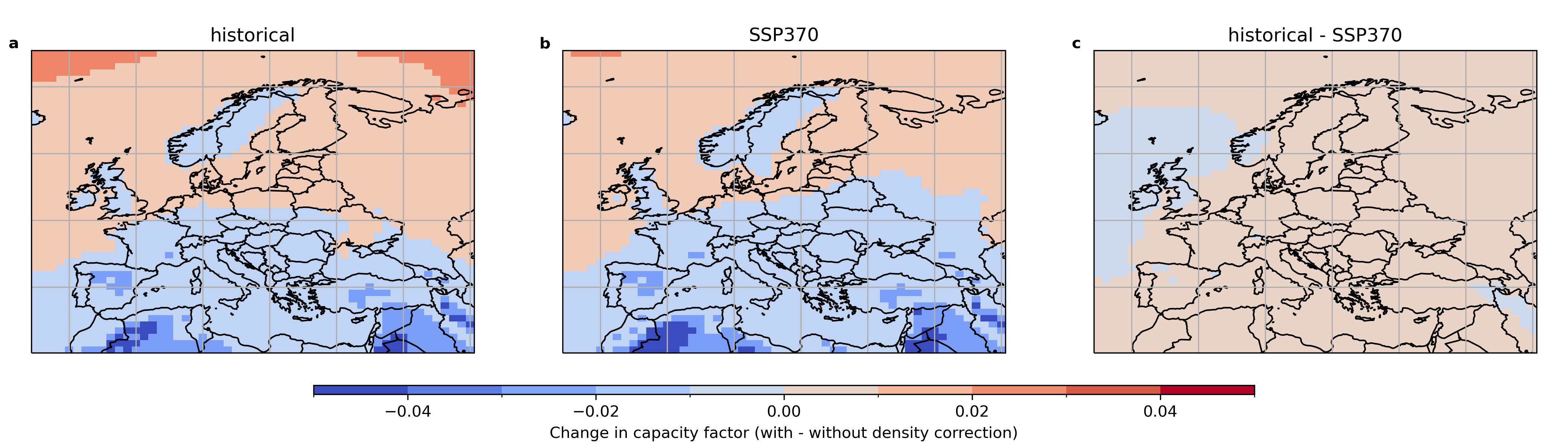}
    \caption{Difference in capacity factors when using variable air density under historical (a) and future (b) climatic conditions. Values are shown for the SWT120\_3600 turbine. Maps show the 20y mean capacity factor computed using variable air density minus the capacity factor using the constant $\rho_\text{IEC}$. Subplot c) shows how the effect of variable air density changes between the historical and future simulation.}
    \label{fig:app_diff_maps_density}
\end{figure}

Fig. \ref{fig:app_mean_maps_density} shows that mean capacity factors are - qualitatively speaking - similar in the historical and SSP3-7.0 scenario independent of whether or not variable air density is included. 
Zooming into the differences, however, Fig. \ref{fig:app_diff_maps_density} reveals that capacity factors are systematically lower in the southern part of Europe, and systematically higher in most of the northern part. 
The magnitude of the effect is about $\Delta CF = \pm 0.01$, which is relevant relative to absolute capacity factors ranging between 0.1 and 0.3 in many onshore locations (Fig. \ref{fig:app_mean_maps_density}).
In other words, ignoring variable air density overestimates CFs in warm Southern Europe and underestimates them in cold Northern Europe by about 0.01 (i.e., between 10\% and 3\% relative change depending on site quality).

While the inclusion of variable air density is thus important to make climate-model based wind energy assessments more physically plausible, variable air density surprisingly has a very similar effect in the historical and SSP3-7.0 scenarios (cf Fig. \ref{fig:app_diff_maps_density} a \& b).
That is, including air density changes only weakly alters climate change impacts on wind capacity factors. 
The sign of the effect, however, is clear in virtually all onshore locations (Fig. \ref{fig:app_diff_maps_density} c): future capacity factors are lower than historical ones when evolving air density is incorporated. 
This finding is physically consistent because higher future temperatures lead to lower air density according to the ideal gas law. Lower densities then translate into lower pseudo wind speeds (eq. \ref{eq:pseudo-wind}) and consequently lower capacity factors. 

\subsection{Variable air density also alters seasonal and daily cycles}

\begin{figure}[ht!]
    \centering
    \includegraphics[width=.8\textwidth]{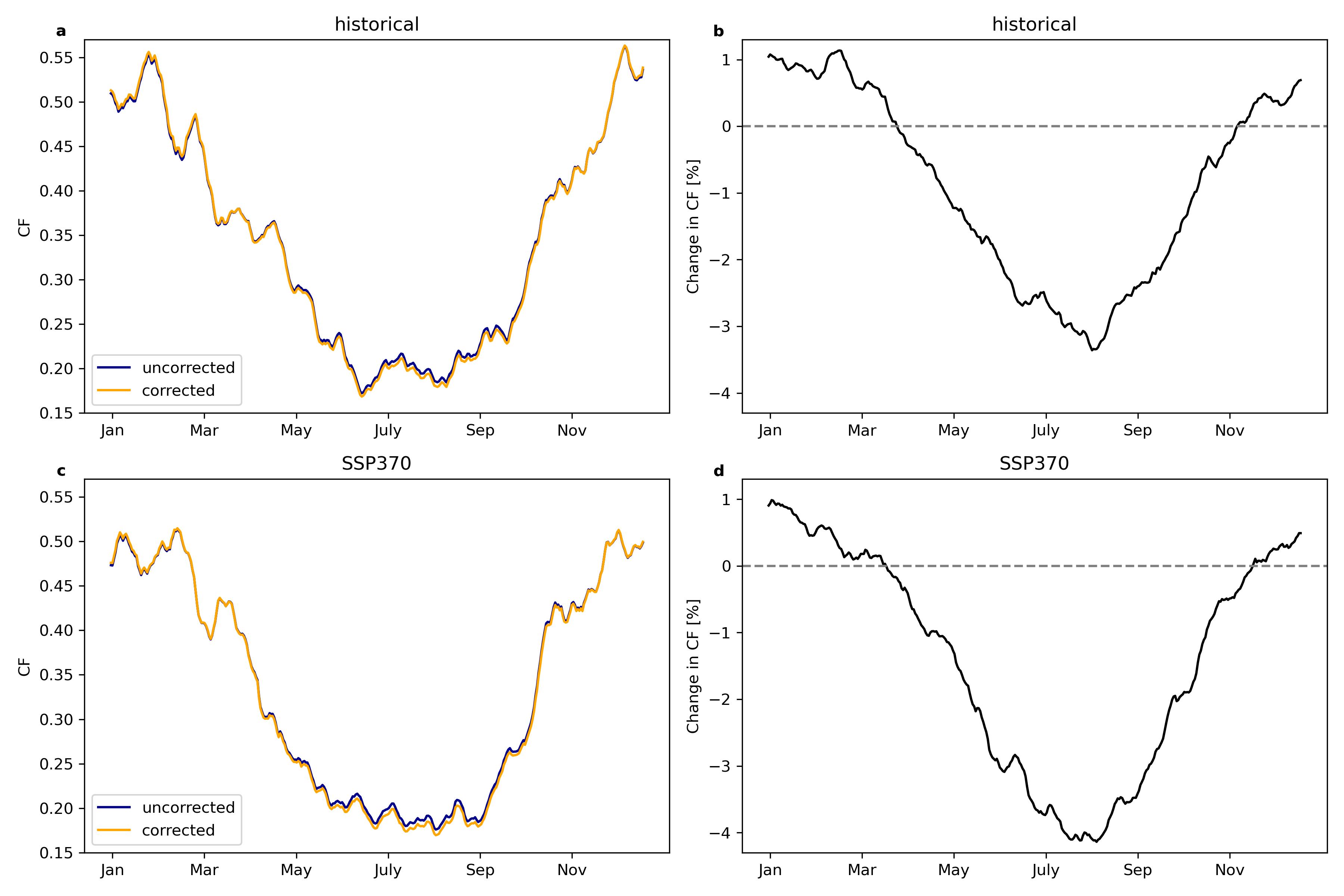}
    \caption{Impact of air density correction on the seasonal cycle in France using the SWT120-3600 turbine. Absolute capacity factors for the historical (a) and SSP3-7.0 scenario (c) are displayed on the left while the subplots on the right display relative CF changes due to the air density correction, seperately for historical (c) and SSP3-7.0 (d). The change in CF is computed as $\frac{\text{corrected} - \text{uncorrected}}{\text{uncorrected}}$ and is displayed in percent.}
    \label{fig:app_seasonal_France_density}
\end{figure}

We use France and the SWT120-3600 turbine as an example that variable air density also alters the seasonal and daily cycle of wind turbine capacity factors. 
Fig. \ref{fig:app_seasonal_France_density} demonstrates systematic alterations in the seasonal cycles, namely higher CFs in winter and lower CFs in summer. 
This pattern is related to higher air density during cold winter conditions and lower air density in warm summer weather. 
It implies an intensification of the seasonal wind CF cycle which already features higher winter CFs due to higher wind speeds. 
While the absolute effect appears weak both under current and future climate (Fig. \ref{fig:app_seasonal_France_density}a,c), the summer reduction reaches 3\% in the historical period and 4\% in the SSP3-7.0 scenario. 
Moreover, the winter CFs increase by around 1\% when accounting for variable air density.

In the context of energy system modeling, these results have two important implications. 
First, wind energy might be able to contribute more to alleviating challenging winter periods (e.g., kalte Dunkelflauten). 
Second, the seasonal cycle of wind generation is underestimated when using fixed air densities and the summer minimum becomes even more pronounced in the future according to the SSP3-7.0 scenario. 
Consequently, optimal technology portfolios will be impacted by including (or ignoring) variable air density.

\begin{figure}[ht!]
    \centering
    \includegraphics[width=.8\textwidth]{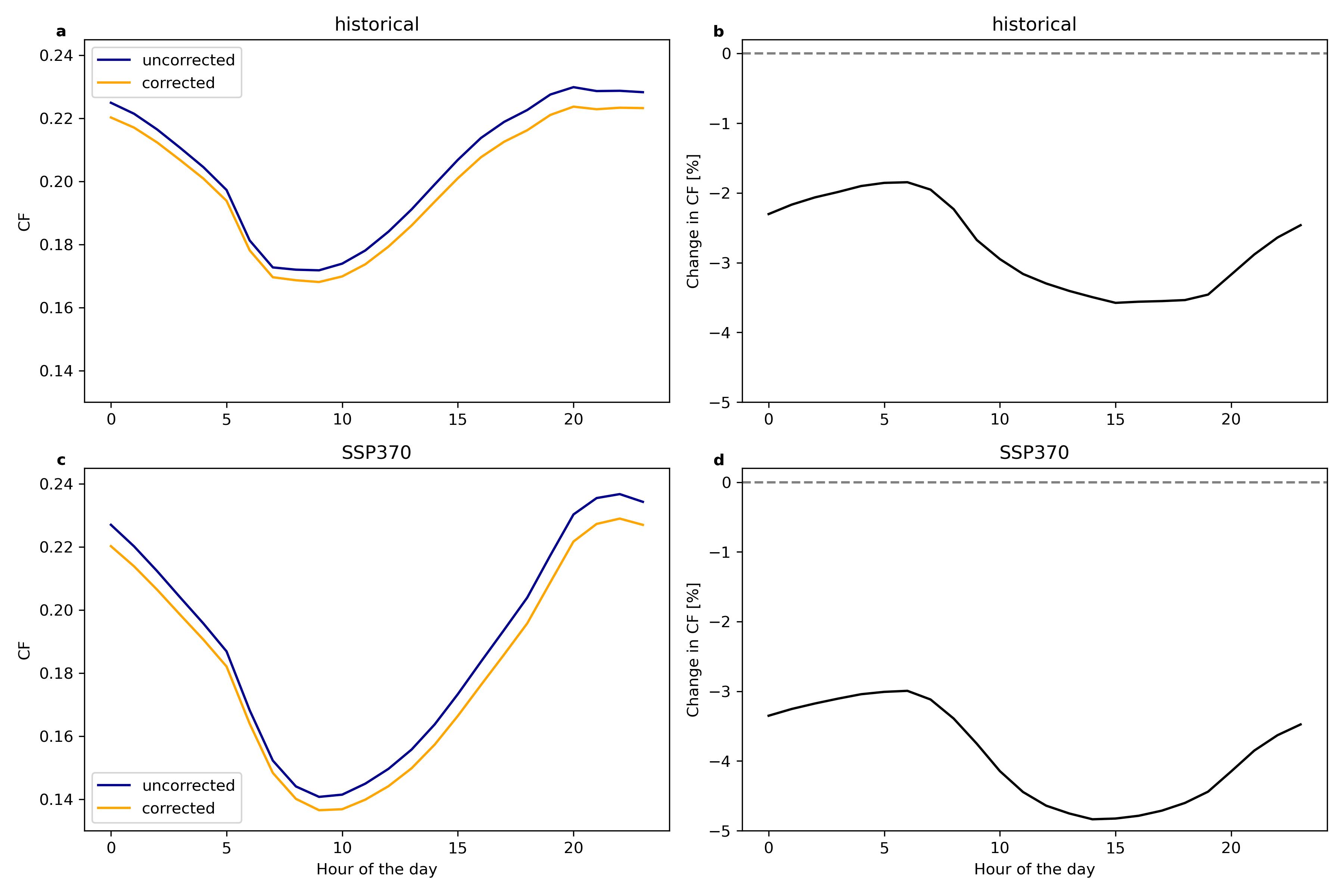}
    \caption{Impact of air density correction on the daily cycle in France using the SWT120-3600 turbine. Absolute capacity factors for the historical (a) and SSP3-7.0 scenario (c) are displayed on the left while the subplots on the right display relative CF changes due to the air density correction, seperately for historical (c) and SSP3-7.0 (d). The change in CF is computed as $\frac{\text{corrected} - \text{uncorrected}}{\text{uncorrected}}$ and is displayed in percent.}
    \label{fig:app_daily_France_density}
\end{figure}

The daily cycle is also impacted, as Fig. \ref{fig:app_daily_France_density} demonstrates for July in France.
CFs are reduced by between -2\% and -3.5\% in the historical, and between -3\% and -5\% in the SSP3-7.0 scenario. 
Night and morning CFs are altered to a lesser degree than midday and afternoon CFs (Fig. \ref{fig:app_daily_France_density} b, d), suggesting a connection to the daily temperature cycle. 
While including variable air density thus has a sizeable effect on July CFs in France, it has less of an impact on the evolution throughout the day, as compared to the changes throughout the year discussed above. 
For instance, the change in CF never becomes positive and always remains in a relatively narrow band. 
Nevertheless, given a pronounced daily cycle in energy consumption, it appears important to account for the effect of variable air density on the ability of wind energy to provide electricity at the right time.

\clearpage

\section{Changes in spatial structure of generation and demand}

\begin{figure}[ht]
    \centering
    \includegraphics[width=.95\textwidth]{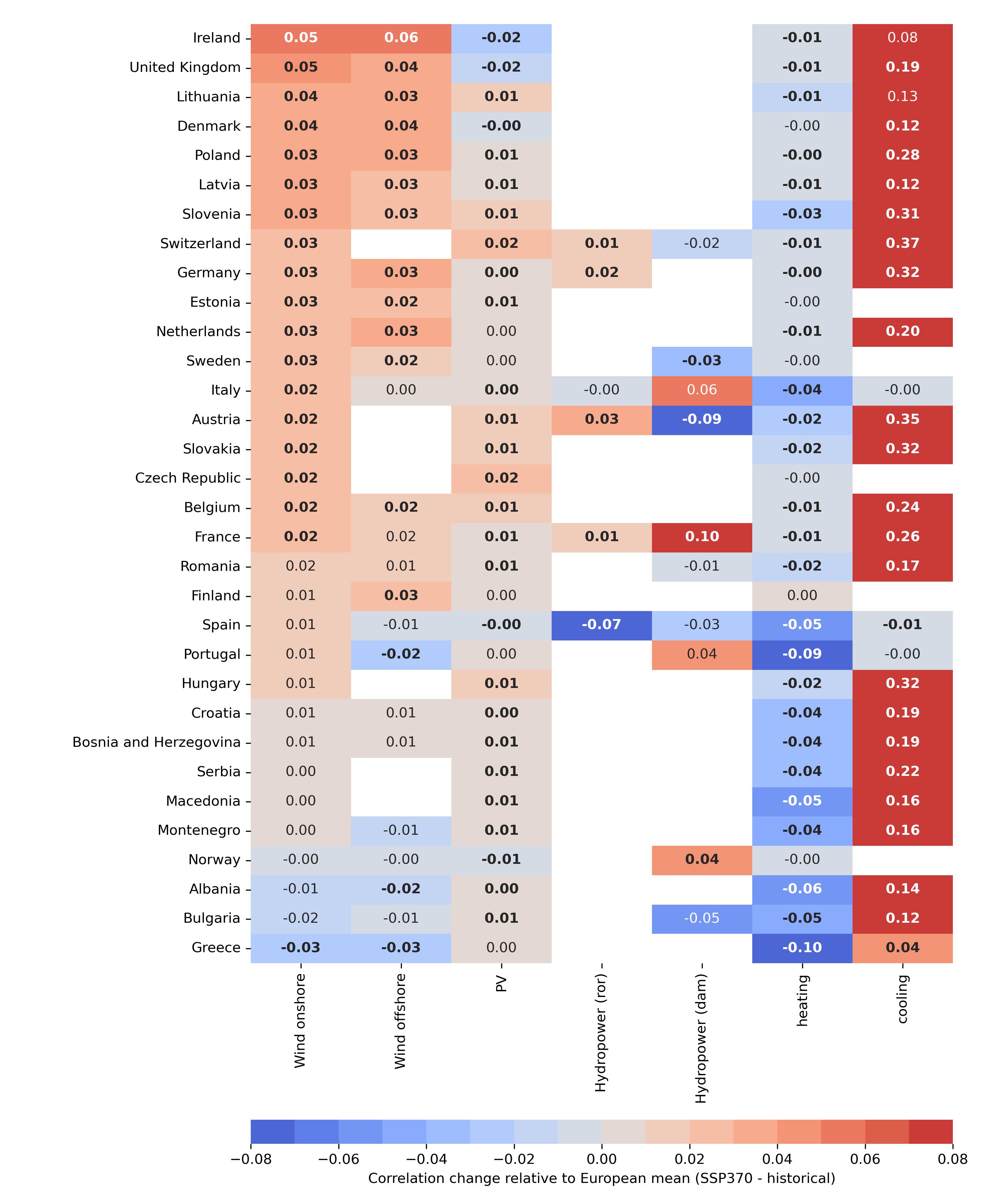}
    \caption{Same as Fig. \ref{fig:correlation_change_subset} but for all countries.}
    \label{fig:SI_correlation_change_all}
\end{figure}

\section{Energy system model}

\begin{figure}[ht]
    \centering
    \includegraphics[width=1.0\textwidth]{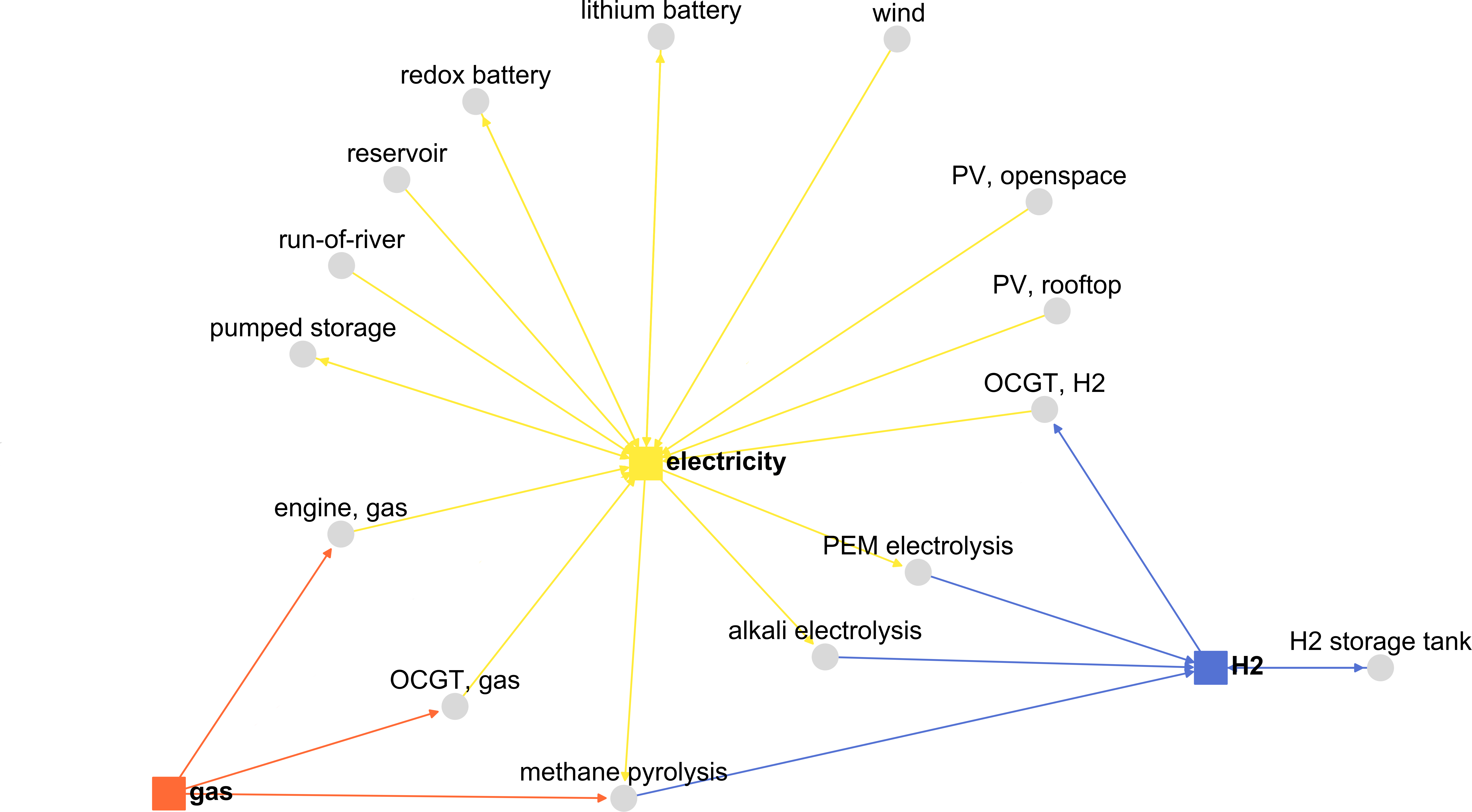}
    \caption{The graph presents the connection between energy technology and carriers in the system model. Energy technologies are depicted as gray circles, and energy carriers as colored squares. Incoming and outgoing edges connect technologies with their input and output carriers, such as electrolyzers using electricity to generate hydrogen. \cite{goke_stabilized_2024}}
    \label{fig:energySystemModel}
\end{figure}

Fig. \ref{fig:energySystemModel} gives an overview of the energy system model. It introduces all considered technologies, depicted as gray circles, and their interaction with energy carriers, depicted as colored squares. Exogenous demand in the model is limited to the carrier electricity which is modeled at an hourly resolution. Hydrogen uses a daily resolution; gas a yearly resolution. 

In the graph, entering edges of technologies refer to their input carriers; outgoing edges relate to outputs. For example, the generation technology electrolysis uses electricity as an input to generate hydrogen. Storage technologies, like pumped storage, have an entering and an outgoing edge to represent charging and discharging. Beyond pumped storage, the model includes batteries for short-term and hydrogen tanks for long-term energy storage. Including long-term storage is critical since it is key for balancing seasonal fluctuations in decarbonized systems. This description here is very similar to the one used in a previous publication using the same model \cite{goke_stabilized_2024}.

\section{Additional Anymod simulations with more climate model realizations}
\begin{figure}[ht]
    \centering
    \includegraphics[width=1.0\linewidth]{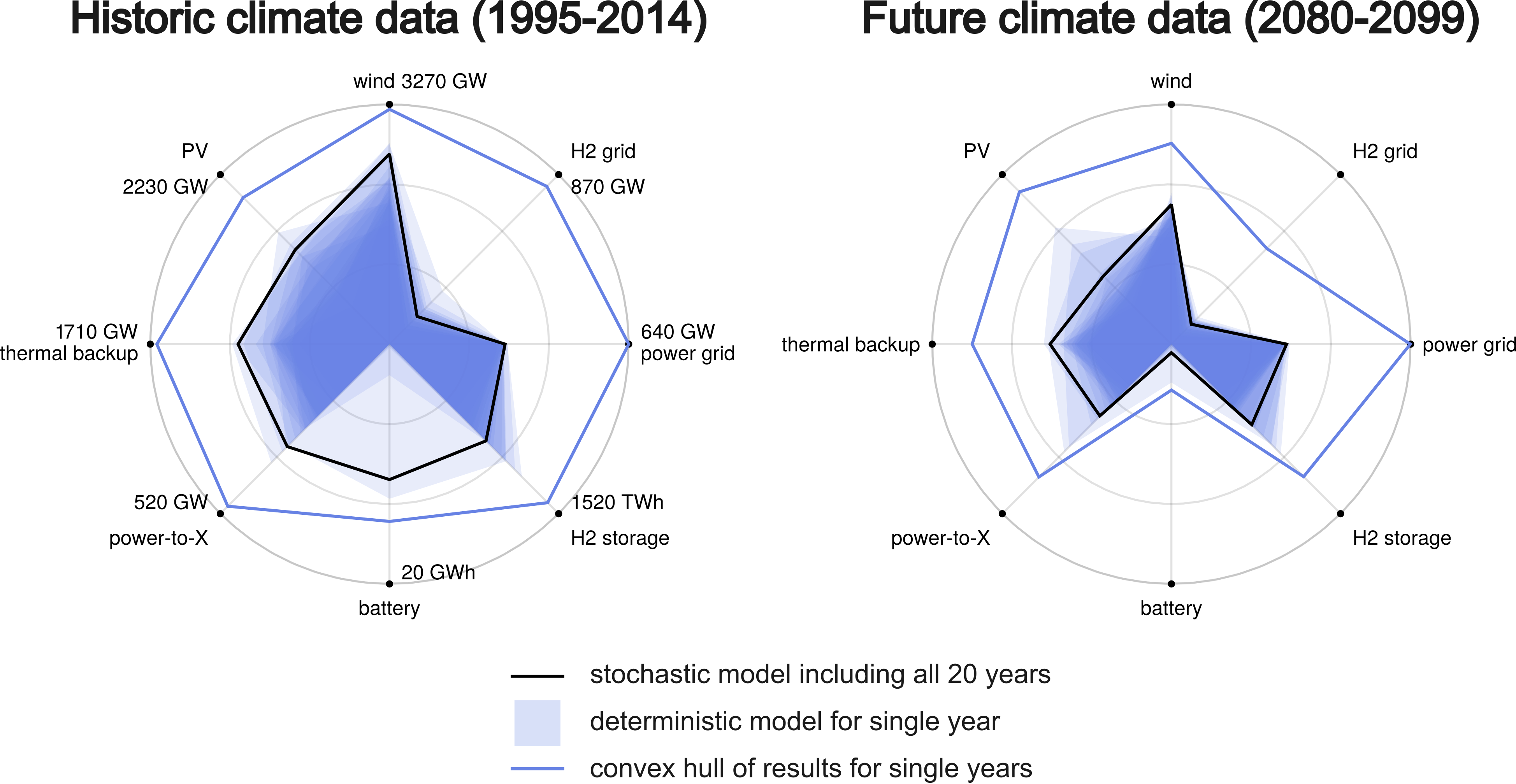}
    \caption{Same as \ref{fig:anymod_optimization} but for CESM realization BB.}
    \label{si_fig:anymod_optimization_bb}
\end{figure}

\begin{figure}[ht]
    \centering
    \includegraphics[width=1.0\linewidth]{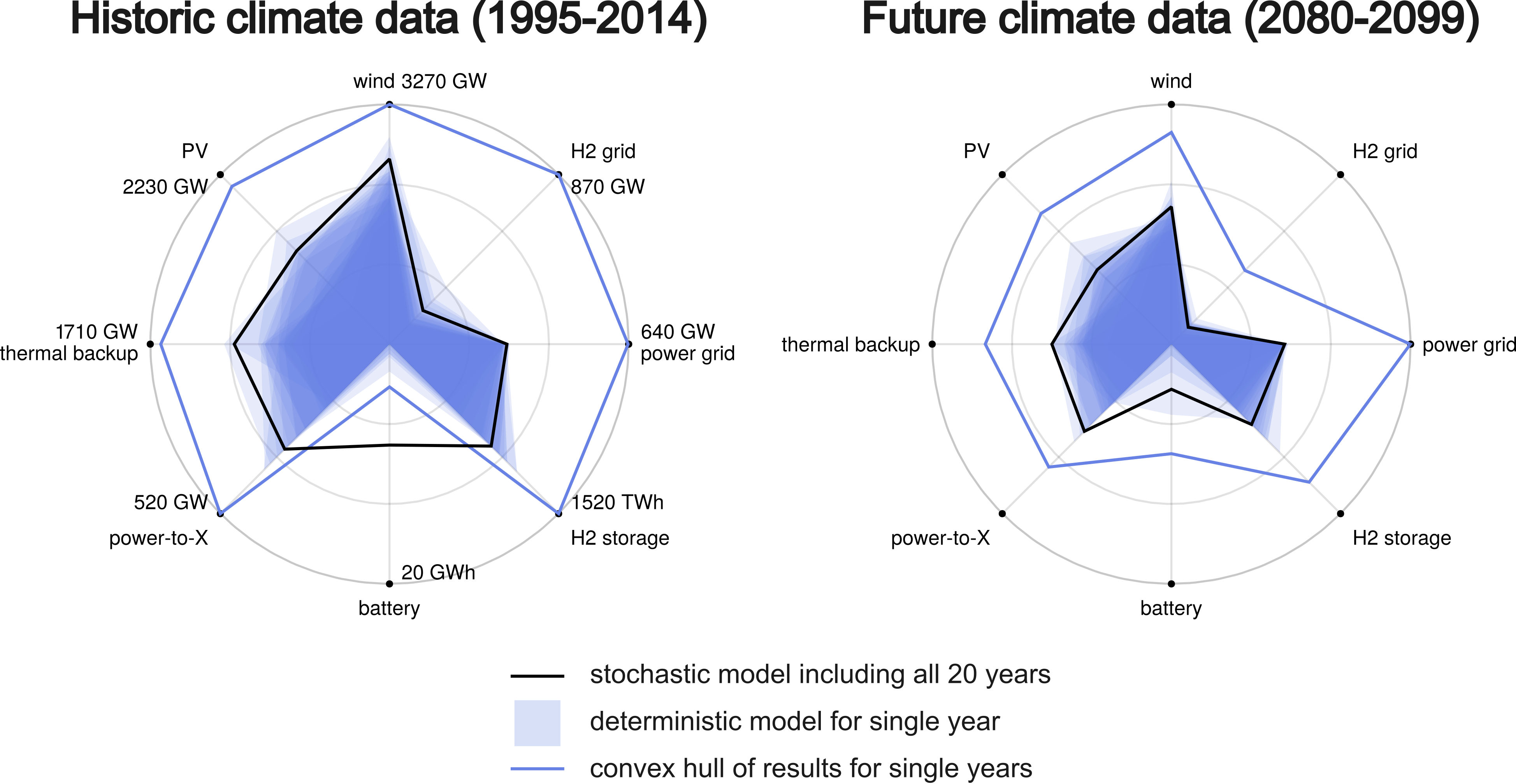}
    \caption{Same as \ref{fig:anymod_optimization} but for CESM realization CC.}
    \label{si_fig:anymod_optimization_cc}
\end{figure}

% uncomment for SI version
%\clearpage
%\bibliography{climate2energy}

\end{appendices}

\end{document}